\documentclass[seceq]{ptptex}
\newcommand{\wlangle}{\left\langle}
\newcommand{\slangle}{\langle}
\newcommand{\wrangle}{\right\rangle_{\!\!t}}
\newcommand{\srangle}{\rangle_t}
\newcommand{\varpitmp}{\xi}

\title{
The adiabatic evolution of orbital parameters in the Kerr spacetime%
}

\author{
Norichika \textsc{Sago}$^1$, 
Takahiro \textsc{Tanaka}$^2$, 
Wataru \textsc{Hikida}$^3$,
Katsuhiko \textsc{Ganz}$^2$ and 
Hiroyuki \textsc{Nakano}$^4$%
}

\inst{
$^1$
Department of Earth and Space Science,~Graduate School of Science,\\
Osaka University, Toyonaka 560-0043, Japan \\
$^2$
Department of Physics,~Graduate School of Science, Kyoto University,\\
Kyoto 606-8502,~Japan \\
$^3$
Yukawa Institute for Theoretical Physics, Kyoto University,\\
Kyoto 606-8502,~Japan \\
$^4$Department of Mathematics and Physics, Graduate School of
Science,\\ Osaka City University, Osaka 558-8585, Japan
}

\abst{
We investigate the adiabatic orbital evolution of a point particle
in the Kerr spacetime due to the emission of gravitational waves.
In the case that the timescale of the orbital evolution is
enough smaller than the typical timescale of orbits,
the evolution of orbits is characterized by the change rates of three
constants of motion, the energy $E$, the azimuthal angular momentum $L$,
and the Carter constant $Q$.
For $E$ and $L$, we can evaluate their change rates from the fluxes
of the energy and the angular momentum at infinity and on the event
horizon according to the balance argument. 
On the other hand, for the Carter constant,
we cannot use the balance argument because we do not know the conserved
current associated with it. 
Recently, Mino proposed a new method of evaluating the averaged 
change rate of the Carter constant by using the radiative field.
In our previous paper
we developed a simplified scheme for practical evaluation of the
evolution of the Carter constant 
based on the Mino's proposal. 
In this paper we describe our scheme in more detail, and 
derive explicit analytic formulae for the change rates of the 
energy, the angular momentum and the Carter constant.
}

\begin{document}

\maketitle

\section{Introduction}
It is believed that supermassive black holes (SMBHs) reside in central
nuclei of many galaxies, and they occasionally capture a stellar mass
compact object (SMCO) which surrounds them.
Gravitational waves from such binary systems with extreme mass
ratios bring us information on the orbits of SMCOs and the spacetime
structure near black holes.
Therefore such systems are considered to be one of the most important
targets of the LISA space-based gravitational wave detector
\cite{LISA}.
In order to detect gravitational waves emitted by extreme mass
ratio inspirals (EMRIs) and to extract physical information from them 
efficiently, we need to predict accurate theoretical waveforms 
in advance.
Our goal along the line of this paper 
is to precisely calculate theoretical waveforms from EMRIs.

To investigate gravitational waves from EMRIs,
our strategy is to adopt the black hole perturbation method:
\cite{Mino:1997bx}
we consider metric perturbations induced by a SMCO in
a black hole spacetime governed by a SMBH. 
We also assume that a SMCO is described by a point particle,  
neglecting its internal structure. 
Under the above approximations, we can calculate the metric perturbation
evaluated at infinity to predict gravitational waveforms.
At the lowest order with respect to the mass ratio, 
we may calculate the metric perturbation 
by approximating the particle's orbit by
a background geodesic.
To step further, 
we consider the orbital shift from the background geodesic by 
taking account of the self-force induced by the particle itself. 

In a Schwarzschild background, we can assume that the 
orbit is in the equatorial plane from the symmetry without loss 
of generality. 
Hence, the orbital velocity can be specified solely by the energy 
and the azimuthal angular momentum of the particle.
Namely, we can evaluate the orbital evolution from 
the change rates of the energy and the angular momentum.
Their averaged change rates 
can be evaluated by using the balance argument; 
the energy and the angular momentum that a particle loses are
equal to the ones that are radiated to the infinity or across the
horizon as gravitational waves because of the conservation laws.
In the limit of a large mass ratio, averaged change rates will 
be sufficient to determine the leading order effects on the 
orbital evolution due to the self-force. 
In this sense the leading order effects can be read from 
the asymptotic 
behavior of the metric perturbation in the Schwarzschild case.

On the other hand, the third constant of motion, i.e., the Carter
constant, is necessary in addition to the energy and the azimuthal 
angular momentum to specify a geodesic in a Kerr background.
However, there is no known conserved current composed of gravitational
waves that is associated with the Carter constant, 
and hence we cannot use the balance argument to evaluate
the change rate of the Carter constant. 
Therefore we have to calculate the self-force acting on a particle
\cite{Ori:1995ed,Ori:1997be}.
When we calculate the self-force, we are faced with the regularization
problem.
Although the formal expression for the regularized 
self-force had been derived
\cite{Mino:1996nk,Quinn:1996am,Detweiler:2002mi},
doing explicit calculation is not so straightforward.

Gal'tsov \cite{Gal'tsov82} proposed a method of calculating
the loss rates of the energy and angular momentum of a particle
by using the radiative part of metric perturbation, which was
introduced earlier by Dirac \cite{Dirac:1938nz}.
The radiative field is defined by half retarded field minus half
advanced one, which is a homogeneous solution of the field equation.
It was shown that the time-averaged loss rates of the energy and
angular momentum evaluated by using the radiative field 
are identical with the results obtained from the balance argument.
Recently, Mino proved that the Gal'tsov's scheme also gives 
the correct averaged change rate of the Carter constant
\cite{Mino:2003yg}.
The Gal'tsov-Mino method has a great advantage that 
we do not need any regularization procedure because
the radiative field is free from divergence from the beginning. 
In Ref.~\citen{Sago:2005gd},
we briefly reported that the formula for the adiabatic evolution
of the Carter constant based on Gal'tsov-Mino method
can be largely simplified. 
In this paper, we explain the derivation of this new formula
in detail.
Applying our new formula, 
we explicitly calculate
the change rate of the Carter constant for orbits with 
small eccentricities and inclinations. 

This paper is organized as follows.
In Sec.~\ref{sec:geodesic},
we give a brief review of the Kerr geometry and the geodesic motion.
Next, we show a practical prescription to calculate the time-averaged
change rates of the constants of motion in Sec.~\ref{sec:COMdot}.
We also derive a simplified expression for the change rate
of the Carter constant.
In Sec.~\ref{sec:example}, applying our prescription,
we calculate the change rates of the constants of motion
and then show the analytic formulae of them
for slightly eccentric and inclined orbits.
Finally we devote Sec.~\ref{sec:summary} to summarize
this paper.
In Appendix~\ref{sec:radiative}, we show the derivation of
the radiative part of metric perturbation.
And we also give short reviews on analytic methods 
of solving the radial Teukolsky equation and 
obtaining the spheroidal harmonics in Appendices~\ref{sec:MST}
and \ref{sec:spheroidal}.

\section{Geodesic motion in the Kerr spacetime} \label{sec:geodesic}
In this section, we give a brief review on geodesics in the 
Kerr geometry.
The metric of the Kerr spacetime in the Boyer-Lindquist
coordinates is 
\begin{eqnarray}
ds^2 &=&
-\left(1-\frac{2Mr}{\Sigma}\right)dt^2
-\frac{4Mar\sin^2\theta}{\Sigma}dtd\varphi
+\frac{\Sigma}{\Delta}dr^2
\nonumber \\ &&
\hspace*{2cm} +\Sigma d\theta^2
+\left(r^2+a^2+\frac{2Ma^2r}{\Sigma}\sin^2\theta\right)
\sin^2\theta d\varphi^2, \label{eq:Kerr}
\end{eqnarray}
where
\[
\Sigma=r^2+a^2\cos^2\theta, \quad
\Delta=r^2-2Mr+a^2.
\]
$M$ and $aM$ are the mass and angular momentum of
the black hole, respectively.
There are two Killing vectors reflecting the stationary
and axisymmetric properties of the Kerr geometry:
\begin{equation}
\xi_{(t)}^{\mu}=(1,0,0,0), \quad
\xi_{(\varphi)}^{\mu}=(0,0,0,1).
\end{equation}
In addition, the Kerr spacetime possesses a Killing tensor,
\begin{equation}
K_{\mu\nu}=2\Sigma l_{(\mu}n_{\nu)}+r^2g_{\mu\nu},
\end{equation}
which satisfies $K_{(\mu\nu;\rho)}=0$, where the parenthese
operating on the indices is the notation for symmetric part of
tensors. Here we have introduced null vectors,
\begin{eqnarray}
l^{\mu} &:=&
\left(\frac{r^2+a^2}{\Delta},1,0,\frac{a}{\Delta} \right), \quad
n^{\mu}:=
\left(\frac{r^2+a^2}{2\Sigma},-\frac{\Delta}{2\Sigma},
0,\frac{a}{2\Sigma}\right),
\nonumber \\
m^{\mu} &:=&
\frac{1}{\sqrt{2}(r+ia\cos\theta)}
\left(ia\sin\theta,0,1,\frac{i}{\sin\theta}\right). 
\end{eqnarray}
We consider a point particle moving in the Kerr geometry:
\[
z^{\alpha}(\tau) =
\left(t_z(\tau),r_z(\tau),\theta_z(\tau),\varphi_z(\tau)\right),
\]
where $\tau$ is the proper time along the orbit.
Here we introduce quantities defined by 
\begin{eqnarray}
\hat{E} &:=&
-u^{\alpha}\xi_{\alpha}^{(t)}=
\left(1-\frac{2Mr_z}{\Sigma}\right)u^t
+\frac{2Mar_z\sin^2\theta_z}{\Sigma}u^{\varphi},
\label{eq:Energy} \\
\hat{L} &:=&
u^{\alpha}\xi_{\alpha}^{(\varphi)}=
-\frac{2Mar_z\sin^2\theta_z}{\Sigma}u^t
+\frac{(r_z^2+a^2)^2-\Delta a^2\sin^2\theta_z}{\Sigma}
\sin^2\theta_z u^\varphi, \label{eq:Momentum} \\
\hat{Q} &:=&
K_{\alpha\beta}u^{\alpha}u^{\beta}
=\frac{(\hat{L}-a\hat{E}\sin^2\theta_z)^2}{\sin^2\theta_z}
+a^2\cos^2\theta_z+\Sigma^2 (u^\theta)^2, \label{eq:Carter}
\end{eqnarray}
where $u^{\alpha}=dz^{\alpha}/d\tau$.
These quantities remain constant as long as the orbit is a geodesic.
$\hat{E}$ and $\hat{L}$ represent the energy and the (azimuthal) angular
momentum per unit mass, respectively.
$\hat{Q}$ is called the Carter constant.
Denoting the mass of a particle by $\mu$, 
the energy, the angular momentum and the Carter constant
of a particle are 
$E\equiv\mu\hat{E}$, $L\equiv\mu\hat{L}$ and $Q\equiv\mu^2\hat{Q}$,
respectively.
Another notation for the Carter constant defined by 
\begin{equation}
C \equiv
Q-(aE-L)^2, \label{eq:Carter2}
\end{equation}
is also convenient since $C$ vanishes for orbits in the 
equatorial plane.
We also use $\hat{C}\equiv C/\mu^2$.

We can specify an orbit of a particle 
by using three constants of motion,
the total energy, angular momentum and Carter constant.
Introducing a new parameter $\lambda$ by 
$d\lambda=d\tau/\Sigma$,
the equations of motion are given as 
\begin{eqnarray}
\frac{dt_z}{d\lambda} &=&
-a(a\hat{E}\sin^2\theta_z-\hat{L})
+\frac{r_z^2+a^2}{\Delta}P(r_z), \label{eq:eom_t}\\
\left(\frac{dr_z}{d\lambda}\right)^2 &=&
R(r_z), \label{eq:eom_r} \\
\left(\frac{d\cos\theta_z}{d\lambda}\right)^2 &=&
\Theta(\cos\theta_z), \label{eq:eom_theta} \\
\frac{d\varphi_z}{d\lambda} &=&
-\left(a\hat{E}-\frac{\hat{L}}{\sin^2\theta_z}\right)
+\frac{a}{\Delta}P(r_z), \label{eq:eom_phi}
\end{eqnarray}
where
\begin{eqnarray}
P(r)&:=&\hat{E}(r^2+a^2)-a\hat{L}, \\
R(r)&:=&[P(r)]^2-\Delta[r^2+(a\hat{E}-\hat{L})^2+\hat{C}], \\
\Theta(\cos\theta)&:=&
\hat{C} - (\hat{C}+a^2(1-\hat{E}^2)+\hat{L}^2)\cos^2\theta
+ a^2(1-\hat{E}^2)\cos^4\theta.
\end{eqnarray}
It should be noted that the equations for $r_z$ and
$\theta_z$ are completely decoupled by using $\lambda$. 
Moreover, $R(r)$ and $\Theta(\cos\theta)$
are quartic functions of $r$ and $\cos\theta$, respectively.

We first consider the radial component of the geodesic equations.
When the radial motion is bounded by the 
minimal and the maximal radii 
$r_{{\rm min}}$ and $ r_{{\rm max}}$,
$r_z(\lambda)$ becomes a periodic function which satisfies 
$r_z(\lambda+\Lambda_r) = r_z(\lambda)$  
with period 
\begin{equation}
\Lambda_r =
2 \int_{r_{{\rm min}}}^{r_{{\rm max}}} \frac{dr}{\sqrt{R(r)}}.
\end{equation}
Therefore, we can expand the radial motion in a Fourier series as
\begin{equation}
r_z(\lambda) = \sum_n \tilde{r}_n e^{-in\Omega_r\lambda}
\,,
\end{equation}
where
\begin{equation}
\Omega_r = {2\pi/\Lambda_r}.
\end{equation}

We can deal with the motion in $\theta$-direction 
in a similar manner.
When the minimum of $\theta$ is given by 
$\theta_{{\rm min}} (\le \pi/2)$, the maximum is 
$\theta_{{\rm max}} = \pi - \theta_{{\rm min}}$
because of the symmetry with respect to the equatorial plane.
As in the case of the radial motion, 
$\cos\theta_z(\lambda)$ becomes a periodic function which satisfies 
$\cos\theta_z(\lambda+\Lambda_\theta) = \cos\theta_z(\lambda)$ 
with period 
\begin{equation}
\Lambda_\theta =
4\int_0^{\cos\theta_{{\rm min}}}
\frac{d(\cos\theta)}{\sqrt{\Theta(\cos\theta)}}.
\end{equation}
We can expand $\cos\theta_z(\lambda)$ in a Fourier series as
\begin{equation}
\cos\theta_z(\lambda) = \sum_n \tilde{z}_n e^{-in\Omega_\theta \lambda}
\,,
\end{equation}
where $\Omega_\theta = {2\pi/\Lambda_\theta}$.

Next, we consider the $t$- and $\varphi$-components of geodesic equations.
Eqs.~(\ref{eq:eom_t}) and (\ref{eq:eom_phi}) can be integrated as
\begin{eqnarray}
t_z(\lambda)&=&t^{(r)}(\lambda)+
   t^{(\theta)}(\lambda)+
   \left\langle {dt_z\over d\lambda}\right\rangle \lambda, \\
\varphi_z(\lambda)&=&\varphi^{(r)}(\lambda)+
   \varphi^{(\theta)}(\lambda)+
   \left\langle {d\varphi_z\over d\lambda}\right\rangle \lambda, 
\end{eqnarray}
where
\begin{eqnarray*}
t^{(r)}(\lambda) &:=&
\int d\lambda \left[ \frac{(r_z^2+a^2)P(r_z)}{\Delta(r_z)}
- \left\langle \frac{(r_z^2+a^2)P(r_z)}{\Delta(r_z)}
  \right\rangle \right], \\
t^{(\theta)}(\lambda) &:=&
-\int d\lambda \left[ a^2 \hat{E}\sin^2\theta_z - a \hat{L}
- \left\langle a^2 \hat{E}\sin^2\theta_z - a \hat{L} \right\rangle
\right], \\
\varphi^{(r)}(\lambda) &:=&
\int d\lambda \left[ \frac{aP(r_z)}{\Delta(r_z)}
- \left\langle \frac{aP(r_z)}{\Delta(r_z)} \right\rangle
\right], \\
\varphi^{(\theta)}(\lambda) &:=&
\int d\lambda \left[ \frac{\hat{L}}{\sin^2\theta_z} -a\hat{E}
- \left\langle \frac{\hat{L}}{\sin^2\theta_z} -a\hat{E} \right\rangle
\right]. \label{eqs:tphi-motion}
\end{eqnarray*}
$\langle \cdots \rangle$ represents the time average along the  
geodesic:
\[
\left\langle F(\lambda) \right\rangle
:=
\lim_{T\to\infty}\frac{1}{2T}\int_{-T}^{T}
d\lambda' \, F(\lambda').
\]
Here, $t^{(r)}(\lambda)$ and $\varphi^{(r)}(\lambda)$ are
periodic functions with period $\Lambda_r$, while
$t^{(\theta)}(\lambda)$ and $\varphi^{(\theta)}(\lambda)$ are 
those with period $\Lambda_\theta$.

\section{The Time Evolution of the Constants of motion}
\label{sec:COMdot}
If the timescale of the orbital evolution due to gravitational
radiation reaction is much longer than the typical dynamical
timescale, we may be able to approximate the particle's motion by 
the geodesic in the background spacetime that is momentarily 
tangential to the orbit 
(osculating geodesic approximation).
Under this assumption, we evaluate the change rates of the constants
of motion at each moment. 
For bound orbits 
we can express the change rates of the constants of motion, 
$I^i=\left\{E,L,Q\right\}$, as 
\begin{eqnarray}
{d I^i\over d\lambda} 
=
\left\langle \frac{dI^i}{d\lambda} \right\rangle 
+ \sum_{(n_r,r_\theta)\not=(0,0)}
      \dot{I}^{i(n_r,n_\theta)}
   \exp\left[ - i(n_r\Omega_r+n_\theta\Omega_\theta) \lambda\right]. 
\label{eq:bareQdot}
\end{eqnarray}
The first term on the right hand side is a time-independent dissipative
contribution due to radiation reaction, while 
the others are oscillating. 
Integrating over a long period, the first term 
becomes dominant. 
In the same spirit in Ref.~\citen{Mino:2003yg},
here we define the 'adiabatic' evolution as an approximation 
which takes account of only the first term. 
Namely, the adiabatic evolution is solely determined by 
the time averaged change rates of the constants
of motion.

Owing to the argument given in Ref.~\citen{Mino:2003yg}, 
we can evaluate the averaged change rates of the constants of 
motion by using
the radiative field of the metric perturbation
\begin{equation}
 \left\langle \frac{dI^i}{d\lambda} \right\rangle =
\lim_{T\to\infty}\frac{1}{2T}\int_{-T}^{T}d\lambda\,\Sigma 
 {\partial I^i\over\partial u^\alpha}
 {f}^{\alpha}[h_{\mu\nu}^{\rm rad}],
\label{eq:meanQdot}
\end{equation}
where $h_{\mu\nu}^{{\rm rad}}$ is the radiative part of the
metric perturbation defined by half retarded field minus half 
advanced field, i.e.,  
$
h_{\mu\nu}^{\rm rad}:=
(h_{\mu\nu}^{\rm ret}-h_{\mu\nu}^{\rm adv})/2. 
$
Radiative field is a solution of source-free vacuum 
Einstein equation. The singular parts contained in both retarded 
and advanced fields cancel out. 
Therefore we can avoid the tedious issue of regularizing 
the self-force. ${f}^{\alpha}$ is a differential operator,
\begin{equation}
f^{\alpha}[h_{\mu\nu}]:=
-\frac{1}{2}(g^{\alpha\beta}+u^{\alpha}u^{\beta})
(h_{\beta\gamma;\delta}+h_{\beta\delta;\gamma}-h_{\gamma\delta;\beta})
u^{\gamma}u^{\delta}.
\end{equation}
This operator with its index lowered reduces to 
\begin{eqnarray}
f_\alpha [h_{\mu\nu}]
&=& g_{\alpha\beta} f^\beta [h_{\mu\nu}] \nonumber \\
&=&
\frac{1}{2}\left( \partial_\alpha h_{\gamma\delta} \right)
u^\gamma u^\delta
-\frac{d}{d\tau}\left( h_{\alpha\gamma}u^\gamma \right)
-\frac{1}{2}u_\alpha \frac{d}{d\tau}
\left( h_{\gamma\delta} u^\gamma u^\delta \right)
+ O(\mu^2),
\end{eqnarray}
ignoring the second order terms.

\subsection{Calculation of $dE/dt$ and $dL/dt$}
From Eq.~(\ref{eq:meanQdot}), we obtain
\begin{eqnarray}
 \left\langle \frac{dE}{d\lambda} \right\rangle &=&
\lim_{T\to\infty}\frac{\mu}{2T}\int_{-T}^{T}d\lambda\,\Sigma 
\left( -\xi_\alpha^{(t)} \right)
 {f}^{\alpha}[h_{\mu\nu}^{\rm rad}] \nonumber \\
&=&
\lim_{T\to\infty}\frac{-\mu}{2T}\int_{-T}^{T}d\lambda
\left[ \frac{\Sigma}{2}\left( \partial_t h_{\gamma\delta}^{{\rm rad}} \right)
       u^\gamma u^\delta
      -\frac{d}{d\lambda}\left(
        h_{t \gamma}^{{\rm rad}}u^\gamma \right)
      +\frac{\hat{E}}{2}\frac{d}{d\lambda}\left(
        h_{\gamma\delta}^{{\rm rad}}u^\gamma u^\delta \right)
\right] \nonumber \\
&=&
\lim_{T\to\infty}\frac{-\mu}{2T}\int_{-T}^{T}d\lambda
\left[ \frac{\Sigma}{2}\left( \partial_t h_{\gamma\delta}^{{\rm rad}} \right)
       u^\gamma u^\delta
\right]. \label{eq:Edot-int}
\end{eqnarray}
In the last equality, 
the total derivative terms are neglected.

Next, we introduce a vector field
$\tilde{u}^\mu (x)$ by~\cite{Sago:2005gd}
\begin{equation}
(\tilde{u}_t, \tilde{u}_r, \tilde{u}_\theta, \tilde{u}_\varphi)
:=
\left( -\hat{E}, \pm\frac{\sqrt{R(r)}}{\Delta(r)},
       \pm\frac{\Theta(\cos\theta)}{\sin\theta}, \hat{L} \right).
\end{equation}
This vector field is a natural extension of the 
four-velocity of a particle. 
In fact, it satisfies
$\tilde{u}_\mu (z(\lambda)) = u_\mu(\lambda)$.
$\tilde{u}_\mu$ depends only on $r$ and $\theta$.
Furthermore, since $\tilde{u}_r$ and $\tilde{u}_\theta$ depend
only on $r$ and $\theta$, respectively, we have the relation
$u_{\mu;\nu} = u_{\nu;\mu}$.

Using this vector field, we can rewrite Eq.~(\ref{eq:Edot-int})
as
\begin{equation}
 \left\langle \frac{dE}{d\lambda} \right\rangle =
\lim_{T\to\infty}\frac{-\mu}{2T}\int_{-T}^{T}d\lambda \left[
\partial_t \left(
\frac{\Sigma}{2}h_{\gamma\delta}^{{\rm rad}}\tilde{u}^\gamma\tilde{u}^\delta
\right) \right]_{x\to z(\lambda)},
\label{eq:dEdl}
\end{equation}
where we used the fact that $\Sigma$ and $\tilde{u}_\mu$
are independent of $t$ (and $\varphi$).

As shown in Appendix~\ref{sec:radiative}
(Eq.~(\ref{hrad})),
the radiative field of metric perturbation is given by
\begin{eqnarray}
h_{\mu\nu}^{{\rm rad}}(x) &=&
\mu \int d\omega \sum_{\ell m} \frac{1}{2i\omega^3} \bigg\{
|N_{s}^{{\rm out}}|^2 {}_s\Pi_{\Lambda,\mu\nu}^{{\rm out}}(x)
\int d\lambda \left[
\Sigma {}_s\bar{\Pi}_{\Lambda,\alpha\beta}^{{\rm out}}(z(\lambda))
u^{\alpha}u^{\beta} \right]
\cr &&
+ \frac{\omega}{k}
|N_{s}^{{\rm down}}|^2 {}_s\Pi_{\Lambda,\mu\nu}^{{\rm down}}(x)
\int d\lambda \left[
\Sigma {}_s\bar{\Pi}_{\Lambda,\alpha\beta}^{{\rm down}}(z(\lambda))
u^{\alpha}u^{\beta} \right] \bigg\}
+ ({\rm c.c.}),
\end{eqnarray}
where $\Lambda=\{\ell m \omega\}$,
$k=\omega-ma/2Mr_+$ and $r_+=M+\sqrt{M^2-a^2}$.
$ {}_s\Pi_{\Lambda,\mu\nu}^{{\rm (out)}}(x) $ 
and $ {}_s\Pi_{\Lambda,\mu\nu}^{{\rm (down)}}(x) $ 
are the out-going and down-going mode solutions 
for $h_{\mu\nu}$, respectively.
$N_s^{{\rm out}}$ and $N_s^{{\rm down}}$
are normalization factors, given by
Eqs.~(\ref{eq:Namp-out}) and (\ref{eq:Namp-down}). 
A bar represents complex conjugation. 
Using this formula, we obtain 
\begin{eqnarray}
\psi^{{\rm rad}}(x) &:=&
\frac{1}{2}\Sigma h_{\gamma\delta}^{{\rm rad}}
\tilde{u}^\gamma \tilde{u}^\delta
\cr &=&
\mu \int d\omega \sum_{\ell m} \frac{1}{4i\omega^3} \bigg[
\phi_\Lambda^{{\rm out}}(x)
\int d\lambda' \bar{\phi}_\Lambda^{{\rm out}}(z(\lambda'))
\cr && \hspace*{2cm}
+ \frac{\omega}{k} \phi_\Lambda^{{\rm down}}(x)
\int d\lambda' \bar{\phi}_\Lambda^{{\rm down}}(z(\lambda'))
\bigg] + ({\rm c.c.}),
\label{eq:huu}
\end{eqnarray}
where
\begin{equation}
\phi_\Lambda^{{\rm (out/down)}}(x) :=
N_s^{{\rm (out/down)}}
\Sigma(x) {}_s\Pi_{\Lambda,\gamma\delta}^{{\rm (out/down)}}(x)
\tilde{u}^\gamma(x) \tilde{u}^\delta(x).
\label{eq:def-phi-out}
\end{equation}
For a bound orbit, we can expand $\phi_\Lambda^{{\rm out}}$
in a Fourier series as:
\begin{equation}
\phi_\Lambda^{{\rm (out/down)}}(z(\lambda)) =
\frac{1}{2\pi} \!
\left\langle\frac{dt_z}{d\lambda}\right\rangle
\!\! \sum_{n_r, n_\theta} \!
\bar{\tilde{Z}}_{\ell m n_r n_\theta}^{{\rm (out/down)}}(\omega)
\exp\left[ i\left\langle\frac{dt_z}{d\lambda}\right\rangle
(\omega - \omega_{mn_r n_\theta})\lambda \right],
\label{eq:phi-out-ft}
\end{equation}
where
\begin{equation}
\omega_{m n_r n_\theta} :=
\left\langle\frac{dt_z}{d\lambda}\right\rangle^{-1}
\left( m \left\langle\frac{d\varphi_z}{d\lambda}\right\rangle
       + n_r \Omega_r + n_\theta \Omega_\theta \right).
\label{eq:disc-omega}
\end{equation}
Substituting Eqs.~(\ref{eq:huu}) and (\ref{eq:phi-out-ft})
into (\ref{eq:dEdl}), we obtain:
\begin{equation}
\wlangle
\frac{dE}{dt}\wrangle =
- \mu^2 \sum_{\ell m n_r n_\theta}
\frac{1}{4\pi\omega_{m n_r n_\theta}^2}
\left( \left|Z_{\ell m n_r n_\theta}^{{\rm out}} \right|^2
+ \frac{\omega_{m n_r n_\theta}}{k_{m n_r n_\theta}}
  \left|Z_{\ell m n_r n_\theta}^{{\rm down}} \right|^2 \right), 
\label{eq:ad-Edot}
\end{equation}
where $k_{m n_r n_\theta}=\omega_{m n_r n_\theta}-ma/2Mr_+$,
$
\slangle F(t) \srangle :=
\lim_{T\to\infty}\frac{1}{2T}\int_{-T}^{T} dt \, F(t)
$, and
\begin{equation}
Z_{\ell m n_r n_\theta}^{{\rm (out/down)}} \equiv
\tilde{Z}_{\ell m n_r n_\theta}^{{\rm (out/down)}}(\omega_{m n_r n_\theta}).
\label{eq:disc-Z}
\end{equation}
In a similar manner, the formula for the loss rate of the angular
momentum is given by 
\begin{equation}
\wlangle\frac{dL}{dt}\wrangle =
- \mu^2 \sum_{\ell m n_r n_\theta}
\frac{m}{4\pi\omega_{m n_r n_\theta}^3} \left(
\left|Z_{\ell m n_r n_\theta}^{{\rm out}} \right|^2
+ \frac{\omega_{m n_r n_\theta}}{k_{m n_r n_\theta}}
  \left|Z_{\ell m n_r n_\theta}^{{\rm down}} \right|^2 \right).
\label{eq:ad-Ldot}
\end{equation}

\subsection{Calculation of $dQ/dt$}
To obtain the change rate of the Carter constant, 
we need to evaluate 
\begin{equation}
 \left\langle \frac{dQ}{d\lambda} \right\rangle =
\lim_{T\to\infty}\frac{\mu^2}{2T}\int_{-T}^{T}d\lambda
2\Sigma K_{\beta}^{\alpha}u^{\beta}
{f}_{\alpha}[h_{\mu\nu}^{\rm rad}].
\end{equation}
Using the vector field $\tilde{u}^{\alpha}(x)$,
which was introduced in the previous subsection,
we obtain 
\begin{eqnarray}
2 K_\beta^\alpha u^\beta f_\alpha &=&
\lim_{x\rightarrow z}\left[
K_\beta^\alpha \tilde{u}^\beta \partial_\alpha
(h_{\gamma\delta}\tilde{u}^\gamma \tilde{u}^\delta)
+2h_{\gamma\delta}\tilde{u}^\beta \tilde{u}^\gamma
(K^\delta_{\beta;\alpha}\tilde{u}^\alpha
 -K^\alpha_\beta \tilde{u}^\delta_{;\alpha})
\right],
\end{eqnarray}
to the first order in perturbation, excluding 
total derivative terms with respect to $\tau$. 
Those total derivative terms do not contribute after
taking a long-time average. 
Furthermore, one can show that 
the second term also vanishes by using 
$K_{(\alpha\beta;\gamma)}=0$ and 
$\tilde{u}_{\alpha;\beta}=\tilde{u}_{\beta;\alpha}$.
After all, we find 
\begin{eqnarray}
 \left\langle \frac{dQ}{d\lambda} \right\rangle &=&
\lim_{T\to\infty}\frac{\mu^2}{2T}\int_{-T}^{T}d\lambda
\left[
2 \Sigma K_\beta^\alpha \tilde{u}^\beta \partial_\alpha
\left( \frac{\psi^{{\rm rad}}(x)}{\Sigma} \right)
\right]_{x\to z(\lambda)} \nonumber \\
& = &
\lim_{T\to\infty}\frac{-\mu^2}{T}\int_{-T}^{T}d\lambda
\nonumber \\ && \times
\left[\left\{
 {P(r)\over \Delta}\left(
  (r^2+a^2)\partial_t+a\partial_{\varphi}\right)
 + {dr_z\over d\lambda}\partial_r
\right\} \psi^{{\rm rad}}(x)\right]_{x\to z(\lambda)}~~.
\label{eq:Qdot-form1}
\end{eqnarray}
To obtain the last term in the last line, 
the term with $\tilde u^\mu\partial_\mu$ was rewritten 
into $\Sigma^{-1}{d/d\lambda}$, and integration by parts  
was applied. 

Substituting Eqs.~(\ref{eq:huu}) and (\ref{eq:phi-out-ft})
into Eq.~(\ref{eq:Qdot-form1}), we obtain:
\begin{eqnarray}
 \left\langle \frac{dQ}{d\lambda} \right\rangle &=&
\lim_{T\to\infty}\frac{-\mu^3}{2T} \int_{-T}^{T}d\lambda
\int d\omega \sum_{\ell m n_r n_\theta}
\frac{1}{2i\omega^3}\delta(\omega-\omega_{m n_r n_\theta})
\cr && \hspace*{-5mm}
\times \bigg[
Z_{\ell m n_r n_\theta}^{{\rm out}}
\bigg\{ \frac{P(r)}{\Delta}
\big( (r^2+a^2)\partial_t + a\partial_\varphi \big)
+\frac{dr_z}{d\lambda}\partial_r \bigg\}
\phi_\Lambda^{{\rm out}}(x)
\cr && \hspace*{-2mm}
+ \frac{\omega}{k}
Z_{\ell m n_r n_\theta}^{{\rm down}}
\bigg\{ \frac{P(r)}{\Delta}
\big( (r^2+a^2)\partial_t + a\partial_\varphi \big)
+\frac{dr_z}{d\lambda}\partial_r \bigg\}
\phi_\Lambda^{{\rm down}}(x) \bigg]_{x\to z(\lambda)}
\nonumber \\ && \hspace*{-2mm} + ({\rm c.c.}).
\label{eq:Qdot-form2}
\end{eqnarray}
Now we focus on the $r$-derivative term in the curly brackets.
Since $\phi_\Lambda^{{\rm out}}$ and $\phi_\Lambda^{{\rm down}}$
depend on $t$ and $\varphi$ only through an exponential function
$e^{-i\omega t + im\varphi}$, we can write 
\begin{eqnarray}
&& \hspace*{-1.5cm}
\phi_\Lambda(z(\lambda))
\delta(\omega-\omega_{m n_r n_\theta})
\nonumber \\ &&
= f(r_z(\lambda), \cos\theta_z(\lambda))
\delta(\omega-\omega_{m n_r n_\theta})
\cr && \hspace*{5mm} \times
\exp\bigg[-i\omega_{m n_r n_\theta} \left(
\left\langle \frac{dt_z}{d\lambda} \right\rangle \lambda
+ t^{(r)}(\lambda) + t^{(\theta)}(\lambda) \right)
\cr && \hspace*{2.5cm}
+ im\left( \left\langle \frac{d\varphi_z}{d\lambda} \right\rangle \lambda
+ \varphi^{(r)}(\lambda) + \varphi^{(\theta)}(\lambda) \right) \bigg]
\nonumber \\ &&
= f(r_z(\lambda), \cos\theta_z(\lambda))
\delta(\omega-\omega_{m n_r n_\theta})
\cr && \hspace*{5mm} \times
\exp\Big[ - in_r\Omega_r\lambda
          - i\omega_{m n_r n_\theta} t^{(r)}(\lambda)
	  + im\varphi^{(r)}(\lambda)
\cr && \hspace*{2.5cm}
          - in_\theta\Omega_\theta\lambda
          - i\omega_{m n_r n_\theta} t^{(\theta)}(\lambda)
	  + im\varphi^{(\theta)}(\lambda)
\Big],
\end{eqnarray}
where $f(r,\cos\theta)$ represents the 
dependence on $r$ and $\cos\theta$ in 
$\phi_\Lambda(x)$. 
$r_z(\lambda)$, $t^{(r)}(\lambda)$ and $\varphi^{(r)}(\lambda)$
are periodic functions with period $\Lambda_r$, while
$\theta_z(\lambda)$, $t^{(\theta)}(\lambda)$ and
$\varphi^{(\theta)}(\lambda)$ are 
those with period $\Lambda_\theta$.
We introduce two different time variables $\lambda_r$
and $\lambda_\theta$. We use them instead of $\lambda$ for functions
with period $\Lambda_r$ and $\Lambda_\theta$.
Then, by using these new variables, we can replace 
the infinitely long time average with a double integral
over a finite region:
\begin{eqnarray}
&& \hspace*{-1cm}
\lim_{T\to\infty}\frac{1}{2T}\int_{-T}^{T} \!\! d\lambda
\delta(\omega-\omega_{m n_r n_\theta})
\frac{dr_z}{d\lambda}
\partial_r \phi_\Lambda(z(\lambda))
\nonumber \\
&&
= \frac{1}{\Lambda_r\Lambda_\theta}
\int_0^{\Lambda_r} \!\!\! d\lambda_r
\int_0^{\Lambda_\theta} \!\!\! d\lambda_\theta
\delta(\omega-\omega_{m n_r n_\theta})
\frac{dr_z}{d\lambda_r} \partial_r\bigg\{
f(r_z(\lambda_r), \cos\theta_z(\lambda_\theta))
\cr && \hspace*{2cm} \times
\exp\bigg[ - in_r\Omega_r\lambda_r
          - i\omega_{m n_r n_\theta} t^{(r)}(\lambda_r)
          + im\varphi^{(r)}(\lambda_r) 
\cr && \hspace*{3.5cm}
          - i n_\theta\Omega_\theta\lambda_\theta
          - i \omega_{m n_r n_\theta} t^{(\theta)}(\lambda_\theta)
          + i m \varphi^{(\theta)}(\lambda_\theta)
\bigg]\bigg\}.
\label{3.22}
\end{eqnarray}
We only need to integrate over one cycle for each of $\lambda_r$ and 
$\lambda_\theta$. Using the relation
\begin{eqnarray*}
&& \hspace*{-1cm}
\frac{d}{d\lambda_r}\left\{
f(r_z(\lambda_r), \cos\theta_z(\lambda_\theta))
\exp[ - in_r\Omega_r\lambda_r
          - i\omega_{m n_r n_\theta} t^{(r)}
          + im\varphi^{(r)} ] 
\right\}
\nonumber \\ &&
= \bigg[ \frac{dt^{(r)}}{d\lambda_r}\partial_t
     + \frac{dr_z}{d\lambda_r}\partial_r
     + \frac{d\varphi^{(r)}}{d\lambda_r}\partial_\varphi
     + \partial_{\lambda_r} \bigg]
\cr && \hspace*{5mm} \times
f(r_z(\lambda_r), \cos\theta_z(\lambda_\theta))
\exp\big[ - in_r\Omega_r\lambda_r
          - i\omega_{m n_r n_\theta} t^{(r)}
          + im\varphi^{(r)} \big],
\end{eqnarray*}
$\lambda_r$-integral in (\ref{3.22}) can be rewritten as 
\begin{eqnarray}
&& \hspace*{-15mm} 
\int_0^{\Lambda_r} \!\!\! d\lambda_r
\frac{dr_z}{d\lambda_r}\partial_r\big\{
f(r_z(\lambda_r), \cos\theta_z(\lambda_\theta))
\cr && \hspace*{1cm} \times
\exp\big[ - i n_r \Omega_r \lambda_r
          - i \omega_{m n_r n_\theta} t^{(r)}
           + i m \varphi^{(r)}
\big] \big\}
\nonumber \\ &=&
\int_0^{\Lambda_r} \!\!\! d\lambda_r \bigg[
- \frac{dt^{(r)}}{d\lambda_r}\partial_t
- \frac{d\varphi^{(r)}}{d\lambda_r}\partial_\varphi
+ i n_r \Omega_r \lambda_r
\bigg]
\nonumber \\
&& \times \big\{
f(r_z(\lambda_r), \cos\theta_z(\lambda_r))
\exp\big[ - i n_r \Omega_r \lambda_r
           - i \omega_{m n_r n_\theta} t^{(r)}
           + i m \varphi^{(r)}
\big] \big\} .
\end{eqnarray}
Eliminating the $r$-derivative term from (\ref{eq:Qdot-form2}) 
by using the above relations,
we obtain 
\begin{eqnarray}
\hspace*{-5mm}\left\langle {dQ\over d\lambda}\right\rangle
& = & 
\lim_{T\to\infty}\frac{-\mu^3}{2T} \int_{-T}^T \!\!\! d\lambda
\int d\omega \!\!\! \sum_{\ell m n_r n_\theta}
\!\! \frac{1}{2i\omega^3} \delta(\omega-\omega_{m n_r n_\theta})
\cr &&
\times \bigg[
Z_{\ell m n_r n_\theta}^{{\rm out}} \left\{
\left\langle {(r^2+a^2)P\over \Delta}\right\rangle \partial_t
+ \left\langle {a P\over \Delta}\right\rangle \partial_\varphi
+ i n_r\Omega_r \right\} \phi_\Lambda^{{\rm out}}(x)
\cr &&
\quad + \frac{\omega}{k}
Z_{\ell m n_r n_\theta}^{{\rm down}} \left\{
\left\langle {(r^2+a^2)P\over \Delta}\right\rangle \partial_t
+ \left\langle {a P\over \Delta}\right\rangle \partial_\varphi
+ i n_r\Omega_r \right\} \phi_\Lambda^{{\rm down}}(x)
\bigg]_{x\to z(\lambda)}
\cr && \quad+ ({\rm c.c.})
\nonumber \\
&=&
- 2 \mu^3 \left\langle \frac{dt_z}{d\lambda} \right\rangle
\!\! \sum_{\ell m n_r n_\theta} \!\!\!
\frac{1}{4\pi\omega_{m n_r n_\theta}^2}
\cr && \hspace*{1cm} \times
\bigg[ - \left\langle {(r^2+a^2)P\over \Delta}\right\rangle
       + \frac{m}{\omega_{m n_r n_\theta}}
         \left\langle {a P\over \Delta}\right\rangle
       + \frac{n_r \Omega_r}{\omega_{m n_r n_\theta}}
\bigg]
\cr && \hspace*{1cm}
\times \left(
|Z_{\ell m n_r n_\theta}^{{\rm out}}|^2
+\frac{\omega_{m n_r n_\theta}}{k_{m n_r n_\theta}}
|Z_{\ell m n_r n_\theta}^{{\rm down}}|^2
\right).
\end{eqnarray}
Here we used Eqs.~(\ref{eq:ad-Edot}) and (\ref{eq:ad-Ldot})
in the last equality. Finally, we obtain:
\begin{eqnarray}
\wlangle {dQ\over dt}\wrangle
& = & 
2\mu\left\langle {(r^2+a^2)P\over \Delta}\right\rangle
 \wlangle{dE\over dt}\wrangle
-2\mu\left\langle {a P\over \Delta}\right\rangle
  \wlangle{dL\over dt}\wrangle
\cr &&
+ \mu^3 \!\!\! \sum_{\ell m n_r n_\theta} \!\!\!
  \frac{n_r \Omega_r}{2 \pi \omega_{m n_r,n_\theta}^3}
\bigg(
|Z_{\ell m n_r n_\theta}^{{\rm out}}|^2
+\frac{\omega_{m n_r n_\theta}}{k_{m n_r n_\theta}}
|Z_{\ell m n_r n_\theta}^{{\rm down}}|^2
\bigg).
\label{eq:ad-Qdot}
\end{eqnarray}

\subsection{Consistency of our formulae in simple cases}
In this subsection, we examine our formulae in a few simple cases.
First, we consider circular orbits. 
We know that a circular orbit remains circular
under radiation 
reaction\cite{Kennefick:1995za}. 
This condition fixes $dQ/dt$ for circular orbits as 
\begin{equation}
\frac{dQ}{dt} =
\frac{2\mu(r^2+a^2)P}{\Delta} \frac{dE}{dt}
- \frac{2\mu aP}{\Delta} \frac{dL}{dt}.
\label{eq:check-cir}
\end{equation}
Since we have $Z_{\ell m n_r n_\theta}^{{\rm out/down}}=0$
for $n_r\ne 0$ in the case of a circular orbit,
the last term in Eq.~(\ref{eq:ad-Qdot}) vanishes. 
Thus Eq.~(\ref{eq:ad-Qdot}) is consistent with the above condition
that a circular orbit remains circular. 

Next, we consider orbits in the equatorial plane.
An orbit in the equatorial plane should not 
leave the plane by symmetry. This can be 
confirmed by rewriting the above formula in terms of $C$. 
From the definition of $\omega_{m n_r n_\theta}$
(\ref{eq:disc-omega}), we obtain the following identity:
\begin{eqnarray*}
&& \hspace*{-1cm}
\mu^2\!\!\sum_{\ell m n_r n_\theta}
\!\! \frac{n_r \Omega_r}{4\pi\omega_{m n_r n_\theta}^3}
\left(
\left|Z_{\ell m n_r n_\theta}^{{\rm out}} \right|^2
+ \frac{\omega_{m n_r n_\theta}}{k_{m n_r n_\theta}}
  \left|Z_{\ell m n_r n_\theta}^{{\rm down}} \right|^2 \right) \\
&=&
\mu^2\!\!\sum_{\ell m n_r n_\theta}
\!\! \frac{1}{4\pi\omega_{m n_r n_\theta}^2} \left(
\left\langle \frac{dt_z}{d\lambda} \right\rangle
-\frac{m}{\omega_{m n_r n_\theta}}
\left\langle \frac{d\varphi_z}{d\lambda} \right\rangle
-\frac{n_\theta \Omega_\theta}{4\pi\omega_{m n_r n_\theta}}
\right)
\cr && \hspace*{2.5cm} \times \left(
\left|Z_{\ell m n_r n_\theta}^{{\rm out}} \right|^2
+ \frac{\omega_{m n_r n_\theta}}{k_{m n_r n_\theta}}
  \left|Z_{\ell m n_r n_\theta}^{{\rm down}} \right|^2 \right) \\
&=&
-\left\langle \frac{dt_z}{d\lambda} \right\rangle
\wlangle \frac{dE}{dt} \wrangle
+\left\langle \frac{d\varphi_z}{d\lambda} \right\rangle
\wlangle \frac{dL}{dt} \wrangle
\cr && \hspace*{1cm}
- \mu^2 \!\! \sum_{\ell m n_r n_\theta}
\!\! \frac{n_\theta \Omega_\theta}{4\pi\omega_{m n_r n_\theta}^3}
\left(
\left|Z_{\ell m n_r n_\theta}^{{\rm out}} \right|^2
+ \frac{\omega_{m n_r n_\theta}}{k_{m n_r n_\theta}}
  \left|Z_{\ell m n_r n_\theta}^{{\rm down}} \right|^2 \right),
\end{eqnarray*}
where we used the the expressions of
$\slangle{dE/dt}\srangle$ and $\slangle{dL/dt}\srangle$ 
given in Eqs.~(\ref{eq:ad-Edot}) and (\ref{eq:ad-Ldot}).
Using this identity, we have
\begin{eqnarray}
\wlangle {dC\over dt}\wrangle
& = &
\wlangle {dQ\over dt}\wrangle
-2(aE-L)\left(
a\wlangle {dE\over dt}\wrangle
-\wlangle {dL\over dt}\wrangle
\right)
\nonumber \\ &=&
- 2\left\langle a^2 E \cos^2\theta_z\right\rangle
   \wlangle{dE\over dt}\wrangle
+ 2\left\langle {L \cot^2\theta_z}\right\rangle
            \wlangle{dL\over dt}\wrangle
\cr &&
-\mu^3 \!\!\!\!\!\sum_{\ell,m,n_r,n_\theta}\!\!\!\!\!
      {n_\theta\Omega_\theta\over 2\pi\omega^3_{m n_r n_\theta}} 
\left(
\left|Z_{\ell m n_r n_\theta}^{{\rm out}} \right|^2
+ \frac{\omega_{m n_r n_\theta}}{k_{m n_r n_\theta}}
  \left|Z_{\ell m n_r n_\theta}^{{\rm down}} \right|^2 \right),
\end{eqnarray}
where we have used the following relations:
\begin{eqnarray*}
\left\langle \frac{dt_z}{d\lambda} \right\rangle &=&
-a(a\hat{E}-\hat{L})
+ \left\langle a^2 \hat{E} \cos^2\theta_z \right\rangle
+ \left\langle \frac{r_z^2+a^2}{\Delta}P \right\rangle, \\
\left\langle \frac{d\varphi_z}{d\lambda} \right\rangle &=&
-a\hat{E} + \hat{L}
+ \left\langle \hat{L} \cot^2\theta_z \right\rangle
+ \left\langle \frac{aP}{\Delta} \right\rangle.
\end{eqnarray*}
From this equation, it is found
that $\slangle dC/dt\srangle=0$ when $\theta=\pi/2$.
Note that we have
$Z_{\ell m n_r n_\theta}^{{\rm out/down}} \ne 0$
only for $n_\theta=0$ in  the case of equatorial orbits.

\section{Application of our formulation to orbits with
small eccentricity and inclination} \label{sec:example}
In this section, as an application of our formulation,
we consider a slightly eccentric orbit with
small inclination from the equatorial plane.
Since, in this case, we can expand an orbit with respect to
the eccentricity and inclination, we can analytically 
calculate the change rates of the constants of motion.

\subsection{Orbits}
Here we define $r_0$ so that the potential in $r$-direction 
$R(r)$ takes its minimum at $r=r_0$:
\begin{equation}
\left. \frac{dR}{dr}\right|_{r=r_0}=0. \label{eq:min-condition}
\end{equation}
We denote the outer turning point by $r_0(1+e)$. 
Namely, 
\begin{equation}
R(r_0(1+e))=0, \label{eq:turn-condition}
\end{equation}
which gives the definition of the eccentricity $e$. 
We also define a parameter $y=C/L^2$, which is 
related to the inclination angle. For orbits in the equatorial 
plane, we have $y=0$.  
Further, we introduce a new parameter $v=\sqrt{M/r_0}$.
For circular orbits 
$v$ represents the orbital velocity at the Newtonian order. 
Hence, we regard $v$ as a parameter whose power indicates 
twice the post-Newtonian (PN) order.

Solving (\ref{eq:min-condition}) and
(\ref{eq:turn-condition}) for $\hat{E}$ and $\hat{L}$, 
they are expressed in terms of 
$e$ and $y$ as 
\begin{eqnarray}
\hat{E} &=&
1-\frac{1}{2} v^2 + \frac{3}{8} v^4 - q v^5
- \left( \frac{1}{2} v^2 - \frac{1}{4} v^4 + 2 q v^5 \right) e^2
+\frac{1}{2} q v^5 y + q v^{5} e^2 y,
\label{eq:E_exp} \\
\hat{L} &=& r_0 v \bigg[
1 + \frac{3}{2} v^2 -3 q v^3
+ \frac{27}{8} v^4 + q^2 v^4
- \frac{15}{2} q v^5
\cr && \hspace*{1cm}
+ \Big( - 1 + \frac{3}{2} v^2 - 6 q v^3
         + \frac{81}{8} v^4 + \frac{7}{2} q^2 v^4
         - \frac{63}{2} q v^5 \Big) e^2
\cr && \hspace*{1cm}
+ \Big( -\frac{1}{2} - \frac{3}{4} v^2 + 3 q v^3 
         - \frac{27}{16} v^4 - \frac{3}{2} q^2 v^4
         + \frac{15}{2} q v^5 \Big) y
\cr && \hspace*{1cm}
+ \Big( \frac{1}{2} - \frac{3}{4} v^2 + 6 q v^3
         - \frac{81}{16} v^4 - \frac{19}{4} q^2 v^4
         + \frac{63}{2} q v^5 \Big) e^2 y
\bigg], 
\label{eq:L_exp}
\end{eqnarray}
where $q:=a/M$.
Hereafter we keep terms up to $O(v^5 e^2 y)$
relative to the leading order.

With the initial condition set to $r_z(\lambda=0)=r_0(1+e)$,
the solution for $r_z(\lambda)$ is obtained in an expansion 
with respect to $e$ as 
\begin{eqnarray}
r_z(\lambda) &=&
r_0[1+er^{(1)}+e^2r^{(2)}],
\end{eqnarray}
where
\begin{eqnarray}
r^{(1)} &=& \cos\Omega_r\lambda, \nonumber \\
r^{(2)} &=&
p^{(1)} (1-\cos\Omega_r\lambda)
+ p^{(2)} (1-\cos 2\Omega_r\lambda), \nonumber \\
\Omega_r &=& r_0 v \bigg[
1 - \frac{3}{2} v^2 + 3 q v^3
- \frac{45}{8} v^4 - \frac{3}{2} q^2 v^4
+ \frac{33}{2} q v^5
\cr && \hspace*{5mm}
- \bigg\{ 1 + \frac{3}{2} v^2 - 6 q v^3
          + \Big( \frac{165}{8} + \frac{9}{2} q^2 \Big) v^4
          - \frac{165}{2} q v^5 \bigg\} e^2
\cr && \hspace*{5mm}
- \Big( \frac{3}{2} q v^3 - 2 q^2 v^4
        + \frac{33}{4} q v^5 \Big) y
- \Big( 3 q v^3 - \frac{27}{4} q^2 v^4 
         + \frac{165}{4} q v^5 \Big) e^2 y
\bigg], \cr
p^{(1)} &=&
 - 1 - v^2 + 2 q v^3 -6 v^4 - q^2 v^4 + 20 q v^5 
- \left( q v^3 - 2 q^2 v^4 + 10 q v^5 \right) y, \cr
p^{(2)} &=&
- \frac{1}{2} - \frac{1}{2} v^2 + q v^3 - 3 v^4
- \frac{1}{2} q^2 v^4 + 10 q v^5
- \Big( \frac{1}{2} q v^3 - q^2 v^4 + 5 q v^5 \Big) y.
\nonumber
\end{eqnarray}

We also compute $\cos\theta_z(\lambda)$ in a series expansion 
in $y$ as
\begin{equation}
\cos\theta_z(\lambda) =
\sqrt{y}[ c_z^{(0)}(\lambda) + y c_z^{(1)}(\lambda)],
\end{equation}
where
\begin{eqnarray*}
c_z^{(0)} &=&
\Big( 1 - \frac{1}{2} q^2 v^4 - \frac{3}{2}q^2 v^4 e^2 \Big)
\sin\Omega_\theta \lambda, \\
c_z^{(1)} &=&
\Big( - \frac{1}{2} + \frac{13}{16} q^2 v^4
       + \frac {39}{16} q^2 v^4 e^2 \Big)
\sin\Omega_\theta \lambda
+ \Big( \frac{1}{16} q^2 v^4 + \frac{3}{16} q^2 v^4 e^2 \Big)
\sin 3\Omega_\theta \lambda, \\
\Omega_\theta &=& r_0 v \bigg[
1 + \frac{3}{2} v^2 - 3 q v^3 + \frac{27}{8} v^4
+ \frac{3}{2} q^2 v^4 - \frac{15}{2} q v^5
\cr && \hspace*{5mm}
+ \Big( - 1  + \frac{3}{2} v^2 -6 q v^3
         + \frac{81}{8} v^4 + \frac{9}{2} q^2 v^4
         - \frac{63}{2} q v^{5} \Big) e^2
\cr && \hspace*{5mm}
+ \Big( \frac{3}{2} q v^3 - \frac{7}{4} q^2 v^4 + \frac{15}{4} q v^5 \Big) y
+ \Big( 3 q v^3 - \frac{9}{2} q^2 v^4 + \frac {63}{4} q v^{5} \Big) e^2 y
\bigg].
\end{eqnarray*}
Here the solution satisfies the condition,
$\cos\theta_z(\lambda=0) = 0$.

Substituting $r_z$ and $\cos\theta_z$
into Eqs.(\ref{eq:eom_t}), (\ref{eq:eom_phi})
and (\ref{eqs:tphi-motion}), we obtain
\begin{eqnarray}
t^{(r)} &=& \frac{r_0 e}{v} \bigg[
\Big\{
( 2 + 4 v^2 - 6 q v^3 + 17 v^4 + 3 q^2 v^4 - 54 q v^5)
\cr && \hspace*{1cm}
+ ( 2 + 6 v^2 - 10 q v^3
  + 33 v^4 + 5 q^2 v^4 - 108 q v^5) e
\cr &&  \hspace*{1cm}
+ ( 3 q v^3 - 4 q^2 v^4 + 27 q v^5) y
\cr &&  \hspace*{1cm}
+ ( 5 q v^3 - 8 q^2 v^4 + 54 q v^5) e y
\Big\} \sin \Omega_r \lambda
\cr && \hspace*{1cm}
+ \Big\{
\Big( \frac{3}{4} + \frac{7}{4} v^2 - \frac{13}{4} q v^3
       + \frac{81}{8} v^4 + \frac{13}{8} q^2 v^4
       - \frac{135}{4} q v^5
\Big) e
\cr && \hspace*{1.5cm}
+ \Big( \frac{13}{8} q v^3 - \frac{5}{2} q^2 v^4
         + \frac{135}{8} q v^5 
\Big) e y
\Big\} \sin 2\Omega_r \lambda
\bigg], \\
t^{(\theta)} &=& q^2 v^3 r_0 y \bigg[
\Big\{
\Big( -\frac{1}{4} + \frac{1}{2} v^2
       - \frac{3}{4} q v^3 + \frac{5}{8} q^2 v^4 + q v^5
\Big)
\cr && \hspace*{1.5cm}
+ \Big( - \frac{1}{4} + \frac{11}{8} v^2 -3 q v^3
         + \frac{1}{2} v^4 + \frac{23}{8} q^2 v^4
         + \frac{9}{2} q v^5
\Big) e^2
\Big\} \sin 2\Omega_\theta \lambda
\bigg], \\
\left\langle {dt_z\over d\lambda}\right\rangle &=&
r_0^2 \Big[
1 + \frac{3}{2} v^2 + \frac{27}{8} v^4 - 3 q v^5
- \Big( \frac{5}{2} + \frac{21}{4} v^2 - 6 q v^3
         + \frac{315}{16} v^4 + 3 q^2 v^4
         - \frac{123}{2} q v^5
\Big) e^2
\cr && \hspace*{5mm}
+ \Big( \frac{1}{2} q^2 v^4 + \frac{3}{2} q v^5
\Big) y
+ \Big( -3 q v^3 + 6 q^2 v^4 - \frac{123}{4} q v^5
\Big) e^2 y \Big], \\
\varphi^{(r)} &=&
q v^3 e \bigg[ \Big\{
( - 2 + 2 q v - 10 v^2 + 18 q v^{3})
+ ( - 2 + 2 q v - 12 v^2 + 24 q v^3) e
\cr && \hspace*{3cm}
- ( q v + 9 q v^3) y 
- ( q v +12 q v^3 ) e y
\Big\} \sin\Omega_r \lambda
\cr && \hspace*{1cm}
+ \Big\{
\Big( - \frac{1}{4} q v + \frac{1}{2} v^2 - \frac{3}{4} q v^3 \Big) e
+ \Big( \frac{1}{8} q v + \frac{3}{8} q v^3 \Big) e y
\Big\} \sin 2\Omega_r \lambda \bigg], \\
\varphi^{(\theta)} &=&
y \bigg[
\Big( - \frac{1}{4} + \frac{3}{8} q^2 v^4 \Big)
+ \frac{9}{8} q^2 v^4 e^2 \bigg]
\sin 2\Omega_\theta \lambda, \\
\left\langle {d\varphi_z\over d\lambda}\right\rangle &=&
r_0 v \bigg[
1 + \frac{3}{2} v^2 - q v^3
+ \frac{27}{8} v^4 - \frac{9}{2} q v^5
- \Big( 1 - \frac{3}{2} v^2 + 2 q v^3
       - \frac{81}{8} v^4 + \frac{27}{2} q v^5
\Big) e^{2}
\cr && \hspace*{1cm}
+ \Big(
\frac{3}{2} q v^3 - q^2 v^4 + \frac{15}{4} q v^5
\Big) y
+ \Big(
3 q v^3 - \frac{9}{4} q^2 v^4 + \frac{63}{4} q v^5
\Big) e^{2} y \bigg].
\end{eqnarray}

\subsection{Calculation of $Z_{\ell m n_r n_\theta}^{{\rm out/down}}$}
In order to obtain the averaged change rates of the energy,
angular momentum and Carter constant, we have to calculate
$Z_{\ell m n_r n_\theta}^{{\rm out/down}}$ defined by
Eq.~(\ref{eq:disc-Z}) with Eq.~(\ref{eq:phi-out-ft}). 
Integrating Eq.~(\ref{eq:def-phi-out}) with respect to
$\lambda$, we obtain
\begin{eqnarray}
&&\hspace*{-1cm}
\hat{Z}_{\Lambda}^{{\rm (out/down)}}
\equiv
\int d\lambda \bar{\phi}_\Lambda^{{\rm (out/down)}} (z(\lambda))
\nonumber \\
&=&
N_s^{{\rm (out/down)}} \int d^4x \sqrt{-g(x)} 
~{}_s\bar{\Pi}_{\Lambda,\alpha\beta}^{{\rm (out/down)}}(x)
\int d\tau
\frac{\tilde{u}^\alpha(x)\tilde{u}^\beta(x)}{\sqrt{-g(x)}}
\delta^{(4)}(x-z(\lambda))
\nonumber \\
&=&
{N_s^{\rm (out/down)}\over \mu} \int d^4x \sqrt{-g(x)}
~{}_s\bar{\Pi}_{\Lambda,\alpha\beta}^{{\rm (out/down)}}(x)
T^{\alpha\beta}(x),
\end{eqnarray}
where 
\begin{equation}
T^{\alpha\beta}(x) =
\mu \int d\tau \frac{1}{\sqrt{-g(x)}}
u^\alpha u^\beta \delta^{(4)}(x-z(\tau)).
\label{ppEM}
\end{equation}
is the energy momentum tensor of a mono-pole particle of mass 
$\mu$. 
Using the relation given in Eq.~(\ref{TT}), 
$\hat{Z}_{\Lambda}^{{\rm (out/down)}}$ can be also expressed 
in the familiar form which appears as an integration over 
the source term in the standard Teukolsky formalism as 
\begin{eqnarray}
\hspace*{-5mm}
\hat{Z}_{\Lambda}^{{\rm (out/down)}} & = &
\frac{N_s^{\rm (out/down)}}{\mu}\bar{\zeta}_s \int d^4x\sqrt{-g(x)} 
~{}_s R_\Lambda^{\rm (in/up)}(r){}_s\bar{Z}_\Lambda(\theta,\varphi)
e^{i\omega t} {}_s\hat{T}(x), 
\label{eq:Zout}
\end{eqnarray}
where ${}_s\hat{T}(x)$ is a projected 
energy momentum tensor defined by 
${}_s \hat{T}:={}_s\tau_{\mu\nu} 
T^{\mu\nu}$ with (\ref{taudef}), and  
$~{}_{s} R_\Lambda^{\rm (in/up)}(r)(={}_{-s}\bar R_\Lambda^{\rm (out/down)}(r))$  and 
${}_s Z_\Lambda(\theta,\varphi)$
are, respectively, the radial mode functions and the spheroidal
harmonics introduced in Appendix~\ref{sec:radiative}.2.

In the following discussion we concentrate on the case with $s=-2$. 
Substituting the explicit forms of 
the energy momentum tensor and the projection operator 
${}_{-2}\tau_{\mu\nu} $, we obtain 
\begin{eqnarray}
\hat{Z}_{\Lambda}^{{\rm (out/down)}}
= 2 N_s^{{\rm (out/down)}}\bar \zeta_s 
\int_{-\infty}^{\infty} dt e^{i\omega t - im\varphi(t)}
{\cal I}_\Lambda^{{\rm (in/up)}}(r(t),\theta(t)),
\end{eqnarray}
with
\begin{eqnarray*}
{\cal I}_\Lambda &=&
\bigg[ R_\Lambda ( A_{nn0}+A_{\bar{m}n0}+A_{\bar{m}\bar{m}0} )
\cr && \hspace*{1cm}
-\frac{dR_\Lambda}{dr}(A_{\bar{m}n1}+A_{\bar{m}\bar{m}1})
+\frac{d^2R_\Lambda}{dr^2}A_{\bar{m}\bar{m}2}
\bigg]_{r=r(t),\theta=\theta(t)}, \\
A_{nn0}&=& \frac{-2}{\sqrt{2\pi}\Delta^2}
C_{nn}\bar z^{2}z\mathcal{L}_{1}^{\dag}
\left\{\bar z^{4}\mathcal{L}_2^{\dagger}(\bar z^{-3} S_\Lambda )\right\},\\
A_{\bar{m}n0}&=& \frac{2}{\sqrt{\pi}\Delta}
C_{\bar{m}n}\bar z^{3}\bigg[
\Big(\frac{iK}{\Delta}+z^{-1}+\bar z^{-1}\Big)
\mathcal{L}^{\dag}_2 S_\Lambda
-\frac{K}{\Delta}(z^{-1}-\bar z^{-1})a\sin\theta S_\Lambda \bigg],\\
A_{\bar{m}\bar{m}0} &=& -\frac{1}{\sqrt{2\pi}}\bar z^{3}z^{-1}
C_{\bar{m}\bar{m}}S_\Lambda\left[-i\left(\frac{K}{\Delta}\right)_{,r}
-\frac{K^2}{\Delta^2}+{2i\over\bar z}\frac{K}{\Delta}\right],\\
A_{\bar{m}n1}&=& \frac{2}{\sqrt{\pi}\Delta}\bar z^{3}
C_{\bar{m}n}\left[\mathcal{L}^{\dag}_2S_\Lambda
+ia\sin\theta(z^{-1}-\bar z^{-1})S_\Lambda\right],\\
A_{\bar{m}\bar{m}1}&=& -\frac{2}{\sqrt{2\pi}}\bar z^{3}z^{-1}
C_{\bar{m}\bar{m}}S_\Lambda\left(i\frac{K}{\Delta}+\bar z^{-1}\right),\\
A_{\bar{m}\bar{m}2}&=&-\frac{1}{\sqrt{2\pi}}\bar z^{3}z^{-1}
C_{\bar{m}\bar{m}}S_\Lambda, \\
C^{\mu\nu}&=&\frac{u^\mu u^\nu}{\Sigma u^t},
\end{eqnarray*}
where $S_\Lambda$ represents ${}_{-2}S_{\Lambda}(\theta)$ defined 
in Appendix.~\ref{sec:radiative}, and 
\begin{eqnarray*}
z &=& r+ia\cos\theta, \\
K &=&
(r^2+a^2)\omega -ma, \\
{\cal L}_s &=&
\partial_\theta + \frac{m}{\sin\theta}
-a\omega\sin\theta + s\cot\theta.
\end{eqnarray*}
Here dagger ($\dag$) means 
an operation that transforms $(m, \omega)$ to $(-m, -\omega)$.
The radial functions and the spheroidal harmonics appearing
in the above equations can be evaluated analytically,
as shown in Appendices~\ref{sec:MST} and \ref{sec:spheroidal}.
For a bound orbit, since 
$e^{-im\varphi(t)}{\cal I}^{\rm (in/up)}_\Lambda(r(t),\theta(t))$ 
is a double periodic function, 
$\hat{Z}_\Lambda^{\rm (out/down)}$ has a discrete spectrum 
as 
\begin{equation}
\hat{Z}_\Lambda^{{\rm out/down}} =
\sum_{n_r, n_\theta} Z_{\ell m n_r n_\theta}^{{\rm out/down}}
\delta(\omega - \omega_{mn_r n_\theta}),
\end{equation}
where the coefficients $Z_{\ell m n_r n_\theta}^{{\rm out/down}}$ are
those already introduced in Eq.~(\ref{eq:disc-Z})
with Eq.~(\ref{eq:phi-out-ft}).
Although we cannot show all the processes explicitly here,
it is straight forward to calculate 
$Z_{\ell m n_r n_\theta}^{{\rm out/down}}$
for each $\omega_{m n_r n_\theta}$
by substituting the analytic expansions of the orbits,
the radial functions and the spheroidal harmonics.

\subsection{Results}
Substituting $Z_{\ell m n_r n_\theta}^{{\rm out/down}}$
obtained by following the scheme explained 
in the preceding subsection into
Eqs.~(\ref{eq:ad-Edot}), (\ref{eq:ad-Ldot}) and (\ref{eq:ad-Qdot}),
we obtain:
\begin{eqnarray}
\wlangle \frac{dE}{dt} \wrangle
&=&
- \frac{32}{5}\left(\frac{\mu}{M}\right)^2 v^{10} 
\cr && \times
\bigg[
1 - \frac{1247}{336} v^2
- \bigg( \frac{73}{12} q - 4 \pi \bigg) v^3
\cr && \hspace*{5mm}
- \bigg( \frac{44711}{9072} - \frac{33}{16} q^2 \bigg) v^4
+ \bigg( \frac{3749}{336} q - \frac{8191}{672} \pi \bigg) v^5
\cr && \hspace*{5mm}
+ \bigg\{ \frac{277}{24} - \frac{4001}{84} v^2
         + \bigg( \frac{3583}{48} \pi - \frac{457}{4} q \bigg) v^3
\cr && \hspace*{1cm}
         + \bigg( 42 q^2 - \frac{1091291}{9072} \bigg) v^4
         + \bigg( \frac{58487}{672} q - \frac{364337}{1344} \pi \bigg) v^5
\bigg\} e^2
\cr && \hspace*{5mm}
+ \bigg( \frac {73}{24} q v^3 - \frac{527}{96} q^2 v^4
         - \frac{3749}{672} q v^5 \bigg) y
\cr && \hspace*{5mm}
+ \bigg( \frac{457}{8} q v^3 - \frac{5407}{48} q^2 v^4
         - \frac{58487}{1344} q v^5 \bigg) e^2 y
\bigg], \\
\wlangle \frac{dL}{dt} \wrangle
&=&
- \frac{32}{5}\left(\frac{\mu^2}{M}\right)  v^{7}
\cr && \times
\bigg[
1 - \frac{1247}{336} v^2
- \bigg( \frac{61}{12} q - 4 \pi \bigg) v^3
\cr && \hspace*{5mm}
- \bigg( \frac{44711}{9072}
- \frac{33}{16} q^2 \bigg) v^4
+ \bigg( \frac{417}{56} q - \frac{8191}{672} \pi \bigg) v^5
\cr && \hspace*{5mm}
+ \bigg\{ \frac{51}{8} - \frac{17203}{672} v^2
+ \bigg( - \frac{781}{12} q + \frac{369}{8} \pi \bigg) v^3
\cr && \hspace*{1cm}
+ \bigg( \frac{929}{32} q^2 - \frac{1680185}{18144} \bigg) v^4
+ \bigg( \frac{1809}{224} q - \frac{48373}{336} \pi \bigg) v^5
\bigg\} e^2
\cr && \hspace*{5mm}
+ \bigg\{ - \frac{1}{2} + \frac{1247}{672} v^2
+ \bigg( \frac{61}{8} q - 2 \pi \bigg) v^3
\cr && \hspace*{1cm}
- \bigg( \frac{213}{32} q^2 - \frac{44711}{18144} \bigg) v^4
- \bigg( \frac{4301}{224} q - \frac{8191}{1344} \pi \bigg) v^5
\bigg\} y
\cr && \hspace*{5mm}
+ \bigg\{ - \frac{51}{16} + \frac{17203}{1344} v^2
+ \bigg( \frac{1513}{16} q - \frac{369}{16} \pi \bigg) v^3
\cr && \hspace*{1cm}
+ \bigg( \frac{1680185}{36288} - \frac{5981}{64} q^2 \bigg) v^4
- \bigg( 168 q - \frac{48373}{672} \pi \bigg) v^5
\bigg\} e^2 y
\bigg], \\
\wlangle \frac{dQ}{dt} \wrangle
&=&
- \frac{64}{5} \mu^3 v^{6}
\cr && \times
\bigg[
1 - q v - \frac{743}{336} v^2
- \bigg( \frac{1637}{336} q - 4 \pi \bigg) v^3
\cr && \hspace*{5mm}
+ \bigg( \frac{439}{48} q^2 - \frac{129193}{18144} - 4 \pi q \bigg) v^4
+ \bigg( \frac{151765}{18144} q - \frac{4159}{672} \pi
        - \frac{33}{16} q^3 \bigg) v^5
\cr && \hspace*{5mm}
+ \bigg\{ \frac{43}{8} - \frac{51}{8} q v - \frac{2425}{224} v^2
       - \bigg( \frac{14869}{224} q - \frac{337}{8} \pi \bigg) v^3
\cr && \hspace*{1cm}
       - \bigg( \frac{453601}{4536} - \frac{3631}{32} q^2
                + \frac{369}{8} \pi q \bigg) v^4
\cr && \hspace*{1cm}
       + \bigg( \frac{141049}{9072} q - \frac{38029}{672} \pi
       - \frac{929}{32} q^3 \bigg) v^5 \bigg\} e^2
\cr && \hspace*{5mm}
+ \bigg\{ \frac{1}{2} q v + \frac{1637}{672} q v^3
       - \bigg( \frac{1355}{96} q^2 - 2 \pi q \bigg) v^4
\cr && \hspace*{1cm}
       - \bigg( \frac{151765}{36288} q - \frac{213}{32} q^3 \bigg) v^5
\bigg\} y
\cr && \hspace*{5mm}
+ \bigg\{ \frac{51}{16} q v + \frac{14869}{448} q v^3
       + \bigg( \frac{369}{16} \pi q - \frac{33257}{192} q^2
         \bigg) v^4
\cr && \hspace*{1cm}
       + \bigg( - \frac{141049}{18144} q + \frac{5981}{64} q^3 \bigg) v^5
  \bigg\} e^2 y
\bigg].
\end{eqnarray}

From the above results we can compute 
\begin{eqnarray}
&& \hspace*{-1cm}
\wlangle \frac{dQ}{dt} \wrangle
- \left\langle \frac{2\mu(r^2+a^2)P}{\Delta}\right\rangle
\wlangle 
   \frac{dE}{dt} \wrangle
+ \left\langle \frac{2\mu aP}{\Delta}\right\rangle
\wlangle 
 \frac{dL}{dt} \wrangle
\nonumber \\ &=&
- \frac{64}{5} \mu^3 v^6 e^2 \bigg[
- \frac{37}{6}
+ \frac{13435}{672} v^2
- \bigg( \frac{1561}{48} \pi - \frac{335}{8} q \bigg) v^3
\cr && \hspace*{2.2cm}
+ \bigg( \frac{625117}{12096} - \frac{337}{32} q^2 \bigg) v^4
+ \bigg( \frac {46827}{448} \pi - \frac{1355}{672} q \bigg) v^5
\cr && \hspace*{2.2cm}
- \bigg( \frac{335}{16} q v^3 - \frac{7559}{192} q^2 v^4
         - \frac{1355}{1344} q v^5
\bigg) y
\bigg].
\end{eqnarray}
The left hand side of this equation vanishes for circular orbits
\cite{Kennefick:1995za}.
In fact, the right hand side vanishes for $e=0$.
When we did not know how to compute $\slangle dQ/dt\srangle$, 
the best guess that we could do for $\slangle dQ/dt\srangle$ 
was to assume that the left hand side vanishes for general 
orbits\cite{Hughes:1999bq}. 
Therefore this combination represents the errors coming from 
this hand-waving working hypothesis. 
We can also compute 
\begin{eqnarray}
\wlangle \frac{dC}{dt} \wrangle &=&
\wlangle \frac{dQ}{dt} \wrangle
-  2(aE - L) \bigg(
a \wlangle \frac{dE}{dt} \wrangle
- \wlangle \frac{dL}{dt} \wrangle
\bigg)
\nonumber \\ &=&
- \frac{64}{5} \mu^3 v^6 y \bigg[
1 - \frac{743}{336} v^2
- \bigg( \frac{85}{8} q - 4 \pi \bigg) v^3
\cr && \hspace*{2.2cm}
- \bigg( \frac{129193}{18144} - \frac{307}{96} q^2 \bigg) v^4
+ \bigg( \frac{2553}{224} q - \frac{4159}{672} \pi \bigg) v^5
\cr && \hspace*{2.2cm}
+ \bigg\{ \frac{43}{8} - \frac{2425}{224} v^2
+ \bigg( \frac{337}{8} \pi - \frac{1793}{16} q \bigg) v^3
\cr && \hspace*{2.7cm}
- \bigg( \frac{453601}{4536} - \frac{7849}{192} q^2 \bigg) v^4
\cr && \hspace*{2.7cm}
+ \bigg( \frac{3421}{224} q - \frac{38029}{672} \pi \bigg) v^5
\bigg\} e^2
\bigg].
\end{eqnarray}
Since $y=0$ (i.e., $C=0$) corresponds to $\theta=\pi/2$, 
$C$ does not 
evolve for equatorial orbits. This is consistent 
with the requirement from the symmetry that an orbit in the
equatorial plane stays in the equatorial plane.

Finally, we consider 
the evolution of an inclination angle $\iota$ defined by 
\cite{Hughes:1999bq},
\begin{equation}
\cos\iota = \frac{L}{\sqrt{L^2+C}}, 
\end{equation}
which, roughly speaking, represents an angle between
the normal vector of an orbital plane and the rotational
axis of the central black hole, 
but this is not the unique definition. 
Although the definition of inclination angle 
can be changed at will to some extent in Kerr case, 
thus defined inclination angle reduces correctly to the 
usual one in the $q=0$ Schwarzschild limit.  
Taking the average of the time derivative of $\cos\iota$,
we obtain
\begin{eqnarray}
\wlangle \frac{d\cos\iota}{dt} \wrangle &=&
\frac{1}{2(L^2+C)^{\frac{3}{2}}}
\bigg( 2 \wlangle \frac{dL}{dt} \wrangle C
       -L \wlangle \frac{dC}{dt} \wrangle
\bigg)
\nonumber \\ &=&
\frac{32\mu^3 v^6}{5L^2(1+y)^{\frac{3}{2}}}
q y \bigg[
\bigg( -\frac{61}{24} v^3 + \frac{13}{96} q v^4
+ \frac{1779}{224} v^5 \bigg)
\cr && \hspace*{2.7cm}
- \bigg( \frac{431}{16} v^3 - \frac{775}{192} q v^4
- \frac{22431}{224} v^5 \bigg) e^2
\bigg].
\end{eqnarray}
Substituting $q=0$ into this equation, we can confirm that 
$\iota$ does not change in the case of Schwarzschild limit,  
which must be so 
because of the spherical symmetry of Schwarzschild spacetime.

\section{Summary} \label{sec:summary}
In this paper, we have considered a scheme to evaluate the change
rates of the orbital parameters of a particle orbiting Kerr black
hole under the adiabatic approximation.
We have adopted the method proposed by Mino \cite{Mino:2003yg},
in which we use the radiative field instead of the retarded
field in order to compute the change rates for ``the 
constants of motion'' due to radiation reaction approximately. 
Based on Mino's method, we have developed a simplified scheme
to evaluate the long term average of the change rates. 
Applying our new scheme, we have performed explicit 
calculations to present analytic formulas of change
rates, $\slangle dE/dt \srangle$,
$\slangle dL/dt \srangle$ and $\slangle dQ/dt \srangle$,
for orbits with small eccentricity and inclination angle.

Here we used the expansions with respect to 
the post-Newtonian order, the eccentricity and the inclination 
angle in evaluating $\slangle dE/dt \srangle$, $\slangle dL/dt \srangle$
and $\slangle dQ/dt \srangle$. 
As a next step therefore we need to examine how large parameter region 
is covered by our formulae with a sufficient accuracy.  
As for the inclination, we recently found a formulation to obtain
the analytic formulae for the change rates 
without assuming a small inclination angle~\cite{Ganz:2005}.
On the other hand, it is almost certain that 
we need numerical calculation for the cases with a 
large eccentricity.
Drasco and Hughes \cite{Drasco:2005kz} developed a numerical code to
calculate the gravitational wave fluxes of energy and azimuthal
angular momentum evaluated at infinity and at the event horizon
for general geodesic orbits.
Fujita and Tagoshi also developed a numerical code based on
an analytic method of solving the radial Teukolsky equation.
By applying such codes to our scheme, we can evaluate
the time-averaged change rate of the Carter constant
for general orbits, although computational cost will not 
be small because we need to take into account a large number 
of frequency modes. 

Once we obtain the change rates of ``the constants of motion'', 
as a next step, we want to use them to trace
the evolution of orbits.
Some strategies to solve the orbital evolution 
taking into account 
the radiation reaction effects were proposed
in Refs.~\citen{Mino:2005an} and \citen{Tanaka:2005ue}.
However, it should be noted that the adiabatic approximation
used here contains only the dissipative part of the self-force
on a particle, and it does not contain the conservative part.
In general, the conservative part also contributes
to the secular evolution of orbits, thought it is not 
the dominant part in the limit $\mu\to 0$. 
Therefore the adiabatic approximation may not be 
sufficient to evaluate the orbital evolution.

Recently, Pound, Poisson and Nickel showed that
the conservative part of the self-force can produce
significant shifts in orbital phases  
in an analogous problem with a charged particle in 
electromagnetism\cite{Pound:2005fs}. 
They suggested that the conservative contribution to
the phase shift is relatively large in weak field, slow motion cases,
while it is suppressed in strong field, rapid motion cases.
Furthermore, there are different types of effects 
higher order in $\mu$ which may produce significant shifts 
in phases.
Therefore it is important to quantify the range of validity 
of the adiabatic approximation for appropriate 
applications of the results obtained in this paper. 
Although it requires computing second order perturbations in 
$\mu$ in order to understand the whole effects which potentially
give phase shifts greater than $O(1)$, 
some of effects can be evaluated by studying 
the first order self-force at each moment without 
averaging over a long period. 
We will come back to this issue in one of our forthcoming 
papers~\cite{Hikida:2005}.

\section*{Acknowledgments}
We would like to thank S.~Drasco, S.~Jhingan,
Y.~Mino, T.~Nakamura, M.~Sasaki and H.~Tagoshi
for invaluable discussions.
NS, WH and HN would like to thank all participants of
the 8th Capra Meeting at the Rutherford Appleton Laboratory
in UK for useful discussions.
This work was supported by Monbukagaku-sho Grant-in-Aid for
Scientific Research of the Japanese Ministry of Education,
Culture, Sports, Science and Technology, Nos. 14047212 and 14047214.
HN and WH are supported by a JSPS Research Fellowship 
for Young Scientists, No.~5919 and No.~1756, respectively.

\begin{appendix}
\section{Radiative solution for the metric perturbation}
\label{sec:radiative}
In this Appendix, we give a brief review on Teukolsky formalism, 
followed by a derivation of the radiative Green function
of the linearized Einstein equations.
This derivation is based on 
Refs.~\citen{Chrzanowski:1975wv,Wald:1978vm} and \citen{Gal'tsov82}.

\subsection{Teukolsky equation}
As a master variable we consider the Teukolsky functions defined by
\begin{eqnarray}
{}_s\Psi &:=& {}_sD^{\mu\nu}h_{\mu\nu} =
\left\{
 \begin{array}{ll}
  -C_{\alpha\beta\gamma\delta}l^{\alpha}m^{\beta}l^{\gamma}m^{\delta},
  & s=2, \\
  -\bar{z}^4C_{\alpha\beta\gamma\delta}
   n^{\alpha}\bar{m}^{\beta}n^{\gamma}\bar{m}^{\delta}, & s=-2,
 \end{array}
\right. 
\label{defPsi}
\end{eqnarray}
where 
\begin{eqnarray}
{}_2 D^{\mu\nu} &=& -{1\over 2z}\bigg[{1\over 2}
   {\cal L}^{\dag}_{-1}{\cal L}^{\dag}_{0}{1\over z}l^\mu l^\nu
   +{\cal D}^2_0 zm^\mu m^\nu 
\cr && \hspace*{1cm}
   -{1\over 2\sqrt{2}}\left( {\cal D}_0{1\over z^2}
    {\cal L}^{\dag}_{-1}z^2+{\cal L}^{\dag}_{-1}{1\over z^2}
    {\cal D}_0 z^2\right)(l^\mu m^\nu +m^\mu l^\nu)\bigg],
\nonumber \\
{}_{-2} D^{\mu\nu} &=& -{1\over 2z}\Biggl[{1\over 2}
   {\cal L}_{-1}{\cal L}_{0}z\bar z^2 n^\mu n^\nu
   +{1\over 4}\Delta^2 {\cal D}^{\dag2}_0{\bar z^2\over z} 
   \bar m^\mu \bar m^\nu 
\cr && \hspace*{5mm}
   +{\Delta^2 \over 4\sqrt{2}}\bigg({\cal D}^{\dag}_0
     {1\over \Delta z^2}
    {\cal L}_{-1}z^2\bar z^2+{\cal L}_{-1}\frac{1}{z^2}
    {\cal D}^{\dag}_0{z^2\bar z^2\over \Delta}\bigg)
    (n^\mu \bar m^\nu +\bar m^\mu n^\nu)\Biggr],
 \label{Ddef}
\end{eqnarray}
$z:=r+ia\cos\theta$,
$\Delta:=r^2-2Mr+a^2$, and $\Sigma:=r^2+a^2\cos^2\theta$.
${\cal D}_n$ and ${\cal L}_s$ are the differential operators defined by 
\begin{eqnarray}
{\cal D}_n &:=&
\partial_r + \frac{(r^2+a^2)}{\Delta}
 \partial_t+\frac{a}{\Delta}\partial_\varphi
+ \frac{2n(r-M)}{\Delta} , \\
{\cal L}_s &:=&
\partial_{\theta} - \frac{i}{\sin\theta}\partial_\varphi
-i a\sin\theta\partial_t + s\cot\theta, 
\end{eqnarray}
and a dagger ($\dag$) acting on an operator 
means transformation 
of $(\partial_t, \partial_\varphi) \to 
(-\partial_t, -\partial_\varphi)$, 
which reduces to the one defined in the main text by 
$(\omega, m) \to (-\omega, -m)$ 
under the assumption of Fourier expansion.
The Teukolsky functions satisfy a separable partial differential
equation~\cite{Teukolsky:1973ha}
\begin{equation}
{}_s{\cal O}~{}_s\Psi = 4\pi\Sigma ~_s\hat{T}, 
 \label{eq:Teukolsky-eq}
\end{equation}
where 
\begin{equation}
{}_s \hat{T}:=~_s\tau_{\mu\nu} T^{\mu\nu},
\end{equation}
and
${}_s{\cal O}$ is the Teukolsky differential operator, 
\begin{eqnarray}
{}_s{\cal O} &:=& {}_s{\cal O}_r + {}_s{\cal O}_\theta,
\end{eqnarray}
with 
\begin{eqnarray}
{}_s{\cal O}_r &:=& 
-\frac{(r^2+a^2)^2}{\Delta}\partial_t^2
+\Delta^{-s}\partial_r(\Delta^{s+1}\partial_r)
-\frac{a^2}{\Delta}\partial_{\varphi}^2
-\frac{4Mar}{\Delta}\partial_t\,\partial_{\varphi}
+\frac{2sa(r-M)}{\Delta}\partial_{\varphi}
\cr &&
+2s\left(\frac{M(r^2-a^2)}{\Delta}-r\right)
\partial_t + s, \cr
{}_s{\cal O}_\theta &:=& 
a^2\sin^2\theta\,\partial_t^2
+\frac{1}{\sin\theta}\,\partial_{\theta}(\sin\theta\,\partial_{\theta})
+\frac{1}{\sin^2\theta}\,\partial_{\varphi}^2
+\frac{2is\cos\theta}{\sin^2\theta}\,\partial_{\varphi}
\cr
&&-2isa\cos\theta\,\partial_t-s^2\cot^2\theta,
\label{calOs}
\end{eqnarray}
and
\begin{eqnarray}
{}_2 \tau_{\mu\nu} & := & {1\over\bar z^4 z}\bigg[
  {1\over\sqrt{2}}\left({\cal L}^{\dag}_{-1}{\bar z^4\over z^2}
  {\cal D}_0+{\cal D}_0{\bar z^4\over z^2}{\cal L}^{\dag}_{-1}
  \right)z^2 (l_\mu m_\nu +m_\mu l_\nu)
\cr && \hspace*{2.5cm}
  -{\cal L}^{\dag}_{-1}\bar z^4{\cal L}^{\dag}_0 {1\over z}l_\mu l_\nu
  -2{\cal D}_0\bar z^4{\cal D}_0 z m_\mu m_\nu\bigg],
 \nonumber \\
 _{-2}\tau_{\mu\nu} & := & -{1\over\bar z^4 z}\Biggl[
  {1\over2\sqrt{2}}\Delta\left({\cal L}_{-1}{\bar z^4\over z^2}
  {\cal D}^{\dag}_{-1}+{\cal D}^{\dag}_{-1}{\bar z^4\over z^2}{\cal L}_{-1}
  \right)\Sigma^2 (n_\mu \bar m_\nu +\bar m_\mu n_\nu)
\cr && \hspace*{2cm}
  +{\cal L}_{-1}\bar z^4{\cal L}_0 \bar z \Sigma n_\mu n_\nu
  +{1\over 2}\Delta^2{\cal D}^{\dag}_0\bar z^4{\cal D}^{\dag}_0 
  {\bar z^2\over z} \bar m_\mu \bar m_\nu\Biggr].
 \label{taudef}
\end{eqnarray}
We consider the following forms of expansions for ${}_s\Psi$
and ${}_s\hat{T}$:
\begin{eqnarray*}
{}_s\Psi &=&
\int_{-\infty}^{\infty}d\omega\sum_{\ell m}
e^{-i\omega t}{}_sX_\Lambda(r){}_sS_\Lambda(\theta)
\frac{e^{im\varphi}}{\sqrt{2\pi}}, \\
4\pi\Sigma~_s\hat{T} &=&
\int_{-\infty}^{\infty}d\omega\sum_{\ell m}
e^{-i\omega t}~_sT_\Lambda(r)~_sS_\Lambda(\theta)
\frac{e^{im\varphi}}{\sqrt{2\pi}},
\end{eqnarray*}
where $\Lambda:=\{lm\omega\}$.
Substituting these expressions into the Teukolsky equation 
(\ref{eq:Teukolsky-eq}),
we obtain equations separated for the radial and angular parts
as
\begin{eqnarray}
\left[\Delta^{-s}\frac{d}{dr}\left(\Delta^{s+1}\frac{d}{dr}\right)
+ \frac{K^2-2is(r-M)K}{\Delta}+4is\omega r
-\lambda \right]{}_sX_\Lambda(r)
&=& {}_sT_\Lambda, \label{eq:radial-Teuk}\\
\bigg[\frac{1}{\sin\theta}\frac{d}{d\theta}
\left(\sin\theta\frac{d}{d\theta}\right)
- a^2\omega^2\sin^2\theta
- \frac{(m+s\cos^2\theta)^2}{\sin^2\theta}
\hspace*{2cm} && \cr 
- 2a\omega s \cos\theta + \lambda + s + 2am\omega
\bigg]{}_sS_\Lambda(\theta)
&=& 0,
\label{eq:spheroid-eq}
\end{eqnarray}
where $K:=(r^2+a^2)\omega -ma$ and
$\lambda := {}_sE_{\ell m}(a\omega) - s(s+1) + a^2\omega^2 - 2am\omega$. 
The eigenvalue ${}_sE_{\ell m}(a\omega)$ is determined 
by solving Eq.(\ref{eq:spheroid-eq}) as an eigenvalue 
problem imposing regular boundary conditions on ${}_sS_\Lambda(\theta)$ 
at $\theta=\pm\pi/2$. Here $\ell$ is an index that labels  
differnt eigen values. We also give a brief review on 
how to solve this equation analytically in Appendix C.

\subsection{Mode functions}
We write mode functions for the Teukolsky equation 
(\ref{eq:Teukolsky-eq}) in the form 
\begin{equation}
{}_s \Omega_\Lambda :=
{}_sR_\Lambda(r)~_s Z_\Lambda(\theta,\varphi) e^{-i\omega t},
 \label{eq:Teuk-mode-fnc}
\end{equation}
where ${}_sR_\Lambda(r)$ is a homogeneous solution of
the radial Teukolsky equation (\ref{eq:radial-Teuk}),
and $_s Z_\Lambda(\theta,\varphi)$ is the 
spheroidal harmonics
\begin{equation}
_sZ_\Lambda(\theta,\varphi) =
\frac{1}{\sqrt{2\pi}}~_sS_\Lambda(\theta) e^{im\varphi},
\end{equation}
normalized as
\begin{equation}
\int_0^\pi d\theta \sin\theta |_sS_\Lambda(\theta)|^2 =1.
\end{equation}
Using the symmetry of the radial equation 
(\ref{eq:radial-Teuk}) under the 
simultaneous operations of the 
complex conjugation and the transformation 
of $(m,\omega)\rightarrow(-m,-\omega)$,
we impose 
\begin{equation}
{}_s R_\Lambda=~_s\bar R^{\dag}_\Lambda,
 \label{Rdag}
\end{equation}
where a dagger ($\dag$) acting on a mode function 
means transformation
of $(\omega, m) \to (-\omega, -m)$. 
In a similar manner, 
by virtue of the symmetries of 
Eq.~(\ref{eq:spheroid-eq}), we arrange 
the spheroidal harmonics to satisfy 
\begin{equation}
{}_s Z_\Lambda =(-1)^m~_{-s}\bar Z_\Lambda^{\dag}. 
\label{eq:symm-spheroid}
\end{equation}
In our later discussions, we also 
need the well-known Teukolsky-Starobinsky identities:
\begin{equation}
 {}_{-s} R_\Lambda =~ _s U _s R_\Lambda,
\qquad
(\mbox{for}~~ |s|=2),  
 \label{Utransform}
\end{equation}
with
\begin{equation}
 {}_{-2} U:={A \over {\cal C}}{\cal D}^4_0,\qquad 
 {}_2 U:={1\over A\bar {\cal C}}\Delta^2
{\cal D}^{\dag 4}_0 \Delta^2,
 \label{Udef}
\end{equation}
where 
\begin{eqnarray}
 {\cal C}& = &[((\lambda+s(s+1))^2
 +4a\omega m -4a^2\omega^2)
   \{(\lambda+s(s+1)-2
    )^2+36a\omega m-36 a^2\omega^2\}\cr
  &&\quad +(2\lambda+2s(s+1)-1)(96a^2\omega^2-48a\omega m)-
   144 a^2\omega^2]^{1/2}+12i\omega M, 
\end{eqnarray}
and $A$ is a factor which depends on how we normalize 
the radial functions.
In this paper, we simply adopt $A=1$. 
(This convention is the one used in Ref.~\citen{Mano:1996gn}). 

Now we discuss how to construct mode functions 
for metric perturbations
from mode functions of the Teukolsky equation. 
The basic idea owes to 
Chrzanowski~\cite{Chrzanowski:1975wv}. 
Here we follow a more rigorous approach taken by 
Wald~\cite{Wald:1978vm}. 
Using the relation (\ref{defPsi}), 
the Teukolsky equation (\ref{eq:Teukolsky-eq}) 
is rewritten as 
\begin{equation}
 {1\over 4\pi\Sigma}{}_s{\cal O}\,_sD^{\mu\nu}
    h_{\mu\nu}={}_s\hat T.
\end{equation}
On the other hand, operating ${}_s\tau_{\alpha\beta}$ on 
the linearlized Einstein equation, which we schematically denote as 
$
 G^{\alpha\beta\mu\nu} h_{\mu\nu}=4\pi T_{\alpha\beta}, 
$
we obtain 
\begin{equation}
{1\over 4\pi}{}_s\tau_{\alpha\beta} G^{\alpha\beta\mu\nu} 
 h_{\mu\nu}={}_s\hat T.
\end{equation}
From the comparison of these equations, we find an identity at the 
operator level:
\begin{equation}
{1\over \Sigma}{}_s{\cal O}\,_sD^{\mu\nu}
  ={}_s\tau_{\alpha\beta} G^{\alpha\beta\mu\nu}.
\label{operatorID}
\end{equation}

Here we define $O^{*\mu\nu}$, 
the adjoint of an operator $O^{\mu\nu}$, so as 
to satisfy 
\begin{equation}
 \int\sqrt{-g} \bar X O^{\mu\nu} Y_{\mu\nu} d^4x
 = \int\sqrt{-g} Y_{\mu\nu}  \overline{ O^{*\mu\nu} X} 
       d^4x, 
\end{equation}
for arbitrary scalar field $X$ and tensor field $Y_{\mu\nu}$. 
The definition of the adjoint operators for 
different types of tensor operators is a straight forward 
generalization of this definition.  
It will be worth noting $\sqrt{-g}\,d^4x=\sin\theta\Sigma \,
dt\,dr\,d\theta\,d\varphi$, 
and 
\begin{eqnarray}
(AB)^*=B^* A^*,\qquad 
{\cal D}_n^*=-\Sigma^{-1}{\cal D}_{-n}^\dag\Sigma,\qquad
{\cal L}_s^*=-\Sigma^{-1}{\cal L}_{1-s}^\dag\Sigma. 
\end{eqnarray}
By taking adjoint of each side in 
Eq.~(\ref{operatorID}), we obtain 
\begin{equation}
{}_sD^{*\mu\nu}
\left(\Sigma^{-1}{}_s{\cal O}\right)^*
  =G^{\alpha\beta\mu\nu} {}_s\tau^*_{\alpha\beta}. 
\end{equation}
Here we used the fact that the linearlized Einstein 
operator $G^{\alpha\beta\mu\nu}$ 
is self-adjoint, i.e., $G^{*\alpha\beta\mu\nu}
=G^{\alpha\beta\mu\nu}$. 
Then, from the definition of ${}_s{\cal O}_r$ and 
${}_s{\cal O}_\theta$ given in Eqs.~(\ref{calOs}), 
it is easy to see that 
\begin{equation}
\left(\Sigma{}^{-1}~_s{\cal O}_r\right)^*=
    \Sigma{}^{-1} {}_{-s}{\cal O}_r, 
\qquad
\left(\Sigma^{-1}~{}_s{\cal O}_\theta\right)^*=
    \Sigma{}^{-1} {}_s{\cal O}_\theta.  
\end{equation}
Therefore we have 
$(\Sigma{}_s{\cal O})^*{}_{-s} R_{\Lambda}\,_s Z_{\Lambda}\,
e^{-i\omega t}=0$,  
whihc means that 
\begin{equation}
G^{\alpha\beta\mu\nu} {}_s\tau^*_{\alpha\beta}\,
 {}_{-s}R_{\Lambda}\,{}_s Z_{\Lambda}\,e^{-i\omega t}=0. 
\end{equation}
Here the explicit form of the 
adjoint operators ${}_s\tau_{\mu\nu}^*$ are  
\begin{eqnarray}
{}_2\tau^*_{\mu\nu} & = &
\bigg[
 {1\over\sqrt{2}}(l_\mu \bar m_\nu +\bar m_\mu l_\nu) 
 {\bar z\over z}\left({\cal D}_0{z^4\over \bar z^2}{\cal L}_2+
 {\cal L}_2 {z^4\over \bar z^2}{\cal D}_0\right)
\cr && \hspace*{1.5cm}
 -l_\mu l_\nu{1\over z\bar z^2}
 {\cal L}_1 z^4{\cal L}_2 -2\bar m_\mu \bar m_\nu {1\over z}
 {\cal D}_0 z^4 {\cal D}_0\bigg]{1\over z^3},
\\
{}_{-2} \tau^*_{\mu\nu} & = &
-\bigg[
 {1\over 2\sqrt{2}}(n_\mu m_\nu +m_\mu n_\nu) 
  z\bar z\left({\cal D}^{\dag}_1{z^4\over \bar z^2}{\cal L}^{\dag}_2+
 {\cal L}^{\dag}_2 {z^4\over \bar z^2}{\cal D}^{\dag}_1\right)\Delta
\cr && \hspace*{1.5cm}
  +n_\mu n_\nu z{\cal L}^{\dag}_1 z^4{\cal L}^{\dag}_2 
  +{1\over 2}m_\mu m_\nu {z\over\bar z^2}
 {\cal D}^{\dag}_0 z^4 {\cal D}^{\dag}_0\Delta^2\bigg]{1\over z^3}. 
\label{taustar}
\end{eqnarray}
Hence, 
\begin{equation}
{}_s\Pi_{\Lambda,\mu\nu} :=
\zeta_s~_s\tau_{\mu\nu}^{*}\,
{}_s\tilde\Omega_\Lambda, 
 \label{Omgtopi}
\end{equation}
with 
\begin{equation}
{}_s\tilde\Omega_\Lambda
 = {}_{-s}R_{\Lambda}\,_s Z_\Lambda \,e^{-i\omega t}, 
\end{equation}
is a complex-valued homogeneous solution of the 
linearized Einstein equations. 
Here $\zeta_s$ is a numerical coefficient which we 
determine so as to satisfy 
\begin{eqnarray}{}_sD^{\mu\nu}{}
\sum_\Lambda ({\cal A}_\Lambda\,_s \Pi_{\Lambda,\mu\nu}
+\overline{{\cal A}_\Lambda\,_s\Pi_{\Lambda,\mu\nu}}) 
=\sum_\Lambda {\cal A}_\Lambda\, {}_s\Omega_{\Lambda}, 
\end{eqnarray}
for any complex-valued amplitude of each mode, ${\cal A}_\Lambda$. 
Here the coplex conjugate term in parentheses
is necessary to make the metric perturbation real. 
Using Eqs.~(\ref{Ddef}) and (\ref{taustar}), we can verify
\begin{eqnarray}
{}_2 D^{\mu\nu}{}_{-2}\tau^*_{\mu\nu} &= &{1\over 4} 
   {\cal L}^{\dag}_{-1} {\cal L}^{\dag}_0 
   {\cal L}^{\dag}_{1} {\cal L}^{\dag}_2, \quad
{}_{-2}D^{\mu\nu}{}_{-2}\tau^*_{\mu\nu}={1\over 16} 
   \Delta^2{\cal D}^{\dag 4}_0\Delta^2,
\nonumber \\
{}_{-2}D^{\mu\nu}{}_{-2}\bar\tau^{*}_{\mu\nu} &=& 0,  \quad
{}_{2}D^{\mu\nu}{}_{2}\tau^*_{\mu\nu}={\cal D}^{4}_0,
\\
{}_{-2}D^{\mu\nu}{}_{2}\tau^*_{\mu\nu} &=&
{1\over 4} {\cal L}_{-1} {\cal L}_0 {\cal L}_{1} {\cal L}_2, \quad
{}_{2}D^{\mu\nu}{}_{2}\bar\tau^{*}_{\mu\nu} = 0. 
\label{Dtau}
\end{eqnarray}
In literature ${}_s\bar\tau^{*\dag}_{\mu\nu}$ is 
used to represent what we denote here by ${}_s\bar\tau^{*}_{\mu\nu}$. 
The difference arises because we use the notation for 
the differential operators without assuming that 
they always act on a single Fourier mode. 
Namely, instead of writing $(-i\omega, im)$, we are using here 
$(\partial_t,\partial_\varphi)$. The complex conjugation 
of the former gives rise a flip of signature, while that 
of the latter does not. 
With the aid of the above relations (\ref{Utransform}) and 
(\ref{Dtau}), we find that the complex conjugate terms vanish
to obtain 
\begin{eqnarray}
 _2 D^{\mu\nu}{}_2\Pi_{\Lambda,\mu\nu} &=&
 \zeta_2 \,{\cal C}\, {}_2\Omega_\Lambda,
 \nonumber \\
 {}_{-2} D^{\mu\nu}{}_{-2}\Pi_{\Lambda,\mu\nu} &=&
 {\zeta_{-2}\, \bar {\cal C}\over 16}{}_{-2}\Omega_\Lambda.
\end{eqnarray}
Thus the normalization constants are fixed as  
\begin{equation}
 \zeta_2={1\over {\cal C}}, \qquad 
 \zeta_{-2}={16\over \bar {\cal C}}.\label{eq:zeta-norm}
\end{equation}

\subsection{Radiative field}
Here we explain a method of constructing radiative field 
for metric perturbations. Radiative field is a homogeneous solution 
of field equations. Hence, once we obtain the radiative field 
for the Teukolsky function, it can be easily transformed into 
that for metric perturbations by using the relations established 
in the preceding subsection. 
We therefore first derive the radiative field 
for the Teukolsky function. 

The retarded Green function of the Teukolsky function 
is defined as a solution of
\begin{equation}
{}_s{\cal O} {}_sG(x,x') =
\frac{\delta^{(4)}(x-x')}{\Delta^s}, 
\end{equation}
with the retarded boundary condition: ${}_sG(x,x') =0$ for 
$t<t'$. 
We write the retarded Green function of the Teukolsky equation 
in the form of Fourier-harmonic expansion as 
\begin{eqnarray}
{}_sG(x,x') &=&
\int \frac{d\omega}{2\pi} \sum_{\ell m}
{}_s g_\Lambda (r,r')
{}_sZ_\Lambda(\theta,\varphi){}_s\bar{Z}_\Lambda(\theta',\varphi')
e^{-i\omega(t-t')}. 
\end{eqnarray}
Then the radial part 
of the Green function 
${}_s g_\Lambda(r,r')$ is given by 
\begin{eqnarray}
{}_s g_\Lambda(r,r') &=&
\frac{1}{W({}_sR_\Lambda^{{\rm in}},{}_sR_\Lambda^{{\rm up}})}
\bigg[
{}_sR_\Lambda^{{\rm up}}(r){}_sR_\Lambda^{{\rm in}}(r')\theta(r-r')
\cr && \hspace*{3cm}
+{}_sR_\Lambda^{{\rm in}}(r){}_sR_\Lambda^{{\rm up}}(r')\theta(r'-r)
\bigg], 
\end{eqnarray}
with the Wronskian defined by 
\begin{eqnarray}
W({}_sR_\Lambda^{{\rm in}},{}_sR_\Lambda^{{\rm up}}) &:=&
\Delta^{s+1}
\bigg[
{}_sR_\Lambda^{{\rm in}}(r)\frac{d}{dr}{}_sR_\Lambda^{{\rm up}}(r)
-{}_sR_\Lambda^{{\rm up}}(r)\frac{d}{dr}{}_sR_\Lambda^{{\rm in}}(r)
\bigg].
\end{eqnarray}

The advanced Green function can be constructed in a similar manner 
just by replacing ``in'' and ``up'' with ``out'' and ``down'', 
respectively. 
Then it is easy to show that 
the radiative Green function has a simple structure which 
does not contain any step function $\theta(r-r')$. 
To show this, let us start with the following expression
for the radial part of the radiative Green function for $r>r'$:
\begin{equation}
{}_sg^{\rm rad}_{\Lambda}(r,r') =
\frac{1}{2}\left[
 \frac{{}_s R^{\rm up}_\Lambda(r){}_s R^{\rm in}_\Lambda(r')}
      {W({}_s R^{\rm in}_\Lambda,{}_s R^{\rm up}_\Lambda)}-
 \frac{{}_s R^{\rm down}_\Lambda(r){}_s R^{\rm out}_\Lambda(r')}
    {W({}_s R^{\rm out}_\Lambda,{}_s R^{\rm down}_\Lambda)}
\right]. 
\label{eq1}
\end{equation}
We rewrite this expression in terms of the 
down-field and the out-field, 
eliminating ${}_s R^{\rm up}_\Lambda(r)$
and ${}_s R^{\rm out}_\Lambda(r')(=\Delta^{-s}
   {}_{-s} \bar R^{\rm in}_\Lambda(r'))$ 
in Eq.~(\ref{eq1}). Hence we expand 
${}_s R^{\rm up}_\Lambda$ and ${}_{s}R^{\rm out}_\Lambda$ as  
\begin{eqnarray}
 {}_s R^{\rm up}_\Lambda 
 & = & \alpha~ {}_s R^{\rm out}_\Lambda
 +\beta~ {}_s R^{\rm down}_\Lambda,\cr
 {}_s R^{\rm out}_\Lambda & = & \gamma~ {}_s R^{\rm up}_\Lambda
  +\delta~ {}_s R^{\rm in}_\Lambda~. 
\label{expansion}
\end{eqnarray}
Taking the Wronskians of both sides of Eqs.~(\ref{expansion})
with appropriate radial functions, 
one can easily obtain 
\begin{eqnarray*}
 && W({}_s R^{\rm up}_\Lambda,{}_s R^{\rm down}_\Lambda)
   =\alpha\, W({}_s R^{\rm out}_\Lambda,{}_s R^{\rm down}_\Lambda),
\qquad
  W({}_s R^{\rm up}_\Lambda,{}_s R^{\rm out}_\Lambda)
   =\beta\, W({}_s R^{\rm down}_\Lambda,{}_s R^{\rm out}_\Lambda), 
\cr
  &&   W({}_s R^{\rm out}_\Lambda,{}_s R^{\rm in}_\Lambda)
   =\gamma\, W({}_s R^{\rm up}_\Lambda,{}_s R^{\rm in}_\Lambda),
\qquad
  W({}_s R^{\rm out}_\Lambda,{}_s R^{\rm up}_\Lambda)
   =\delta\, W({}_s R^{\rm in}_\Lambda,{}_s R^{\rm up}_\Lambda). 
\end{eqnarray*}
Substituting these relations, the expression (\ref{eq1}) 
reduces to
\begin{eqnarray}
_sg^{\rm rad}_\Lambda(r,r')&=& {\Delta^{-s}(r')\over
    2 W({}_s R^{\rm in}_\Lambda,{}_s R^{\rm up}_\Lambda)
    W({}_s R^{\rm out}_\Lambda,{}_s R^{\rm down}_\Lambda) }\cr
&&\quad   \times \Bigl[
    W({}_s R^{\rm out}_\Lambda,{}_s R^{\rm in}_\Lambda) 
     \,_sR^{\rm down}_{\Lambda}(r) 
     \,_{-s}\bar R^{\rm down}_{\Lambda}(r')\cr
  &&\hspace*{2cm} 
     +W({}_s R^{\rm up}_\Lambda,{}_s R^{\rm down}_\Lambda) 
    \,_s R^{\rm out}_{\Lambda}(r) 
    \,_{-s}\bar{R}^{\rm out}_{\Lambda}(r') 
    \Bigr].  \label{eq2}
\end{eqnarray}
We can do an analogous reduction for $r<r'$, and the result 
turns out to be the same as that 
for $r>r'$. Namely, the step functions 
which was present in the retarded and the advanced Green functions 
do not appear in the radiative Green function. 
This is consistent with the fact 
that the radiative field is a source-free 
homogeneous solution. 

Since the radiative field is a homogeneous solution, 
we can use the method for reconstruction of metric 
perturbation explained in the preceding subsection. 
When we consider the metric perturbation by a point mass,
the energy-momentum tensor is given by (\ref{ppEM}). 
In this case it is easy to verify that the radiative field 
of the metric perturbations is given by 
\begin{eqnarray}
h_{\mu\nu}^{{\rm rad}}(x) &=&
\mu \! \int \!\! 
d\omega \sum_{\ell m} 
\bigg\{
{\cal N}_{s}^{\rm out} 
{}_s\Pi_{\Lambda,\mu\nu}^{{\rm out}}(x)
\int \!\! d\tau \Big[
{}_s\bar{\Pi}_{\Lambda,\alpha\beta}^{{\rm out}}(z(\lambda))
u^{\alpha}u^{\beta} \Big]
\cr &&
+ {\cal N}_{s}^{\rm down} {}_s\Pi_{\Lambda,\mu\nu}^{{\rm down}}(x)
\int \!\! d\tau \Big[
{}_s\bar{\Pi}_{\Lambda,\alpha\beta}^{{\rm down}}(z(\lambda))
u^{\alpha}u^{\beta}
\Big] \bigg\}
+ ({\rm c.c.}), 
\label{eq:rad-field}
\end{eqnarray}
with 
\begin{eqnarray}
 {\cal N}_{s}^{\rm out} 
  & = & { W({}_s R^{\rm up}_\Lambda,{}_s R^{\rm down}_\Lambda)
    \over \bar\zeta_s 
    W({}_s R^{\rm in}_\Lambda,{}_s R^{\rm up}_\Lambda)
    W({}_s R^{\rm out}_\Lambda,{}_s R^{\rm down}_\Lambda) },\cr
 {\cal N}_{s}^{\rm down} 
  & = & { W({}_s R^{\rm out}_\Lambda,{}_s R^{\rm in}_\Lambda)
    \over \bar\zeta_s 
    W({}_s R^{\rm in}_\Lambda,{}_s R^{\rm up}_\Lambda)
    W({}_s R^{\rm out}_\Lambda,{}_s R^{\rm down}_\Lambda) }. 
\end{eqnarray}
In fact, if we apply ${}_s{\cal D}^{\mu\nu}$, we correctly 
recover
${}_s \Psi^{\rm rad}(x)=4\pi\int 
G^{\rm rad}(x,x') \Sigma(x')$
$
\Delta^s(x') {}_s\hat T(x') d^4 x$. 
To show this, we also used
\begin{eqnarray}
\bar\zeta_s \int \sqrt{-g}
   {}_s\bar{\tilde{\Omega}}_{\Lambda} 
       \,{}_s\hat T d^4x
&=&\bar\zeta_s\int \sqrt{-g} 
  \overline{\left({}_s\tau^{*}_{\mu\nu}\, 
   {}_s\tilde\Omega_{\Lambda}\right)} T^{\mu\nu} d^4x \cr
&=&\mu\int d\tau\,
    u^{\mu}  u^{\nu} 
    \bar\Pi_{\Lambda,\mu\nu}(z(\tau)). 
\label{TT}
\end{eqnarray}

It is more convenient to rewrite ${\cal N}_s$ written 
in terms of Wronskians by using the coefficients 
in the asymptotic forms of radial functions. 
The radial functions take the asymptotic forms,
\begin{eqnarray}
{}_s R_\Lambda^{{\rm in}} &:=&
\left\{
\begin{array}{ll}
 {}_sB_{\Lambda}^{{\rm inc}}\displaystyle r^{-1} e^{-i\omega r^{*}}
 +{}_sB_{\Lambda}^{{\rm ref}} r^{-2s-1} e^{i\omega r^{*}}, \hskip1cm
&
 {\rm for}~~r^{*}\rightarrow\infty,
\\
 \displaystyle
 {}_sB_{\Lambda}^{{\rm trans}}\Delta^{-s}e^{-ik r^{*}},
&
 {\rm for}~~r^{*}\rightarrow -\infty,
\end{array}
\right. \label{eq:asymptotic-in} \\
{}_s R_\Lambda^{{\rm up}} &:=&
\left\{
\begin{array}{ll}
 \displaystyle
  {}_sC_{\Lambda}^{{\rm trans}}r^{-2s-1} e^{i\omega r^{*}}, \hskip3cm
\hspace*{7mm}
&
  {\rm for}~~r^{*}\rightarrow\infty, 
\\
 \displaystyle
 {}_sC_{\Lambda}^{{\rm up}}e^{ik r^{*}}
 + {}_sC_{\Lambda}^{{\rm ref}}\Delta^{-s}e^{-ik r^{*}},
&
 {\rm for}~~r^{*}\rightarrow -\infty, 
\end{array}
\right.
\label{eq:asymptotic-up}
\end{eqnarray}
where $r^*$ is the tortoise coordinate defined by
$dr^*/dr=(r^2+a^2)/\Delta$.
Using the relations 
${}_s R^{\rm out}_\Lambda=
\Delta^{-s}{}_{-s} \bar R^{\rm out}_\Lambda$
and
${}_s R^{\rm down}_\Lambda=
\Delta^{-s}{}_{-s} \bar R^{\rm up}_\Lambda$, 
we can describe the asymptotic forms
of out- and down- fields with the same coefficients
that appear in Eqs.~(\ref{eq:asymptotic-in}) and
(\ref{eq:asymptotic-up}).
Then, the Wronskians that we need to evaluate are 
\begin{eqnarray}
    W({}_s R^{\rm in}_\Lambda
,{}_s R^{\rm up}_\Lambda) 
& = & 
  2i\omega\,  {}_{s}B_\Lambda^{\rm inc} \,
      {}_{s}C_\Lambda^{\rm trans}, 
\cr 
    W({}_s R^{\rm out}_\Lambda,{}_s R^{\rm down}_\Lambda) 
& = & 
  -2i\omega\,  {}_{-s}\bar C_\Lambda^{\rm trans} \,
      {}_{-s}\bar B_\Lambda^{\rm inc}, \cr 
    W({}_s R^{\rm out}_\Lambda,{}_s R^{\rm in}_\Lambda) 
& = & 
  -4ikM r_+\kappa_s \,  {}_{s} B_\Lambda^{\rm trans} \,
      {}_{-s}\bar B_\Lambda^{\rm trans}, \cr 
    W({}_s R^{\rm up}_\Lambda,{}_s R^{\rm down}_\Lambda) 
& = & 
  -2i\omega\,  {}_{-s}\bar C_\Lambda^{\rm trans} \,
      {}_{s} C_\Lambda^{\rm trans}, 
\end{eqnarray}
where $\displaystyle\kappa_s:=1-{is(r_+-M)/ 2kMr_{+}}$.
The coefficients with $(-s)$-spin can be erased by using 
the Teukolsky-Starobinsky identities~(\ref{Utransform}). 
Substituting the asymptotic forms (\ref{eq:asymptotic-in}) and
(\ref{eq:asymptotic-up}) into Eqs.~(\ref{Utransform}), we obtain
\begin{eqnarray}
 {}_{-2}B_\Lambda^{\rm inc} &= 
  &{{\cal C}\over(2\omega)^4}{}_{2}B_\Lambda^{\rm inc}, \qquad
 {}_{-2}B_\Lambda^{\rm trans}
   =  \left({1\over 4M r_{+} k}\right)^4
  {{\cal C}
   \over \kappa_{-2}\kappa_{-1}\kappa_1}{}_{2}B_\Lambda^{\rm trans}, 
 \nonumber \\
  {}_{-2}C_\Lambda^{\rm trans} 
 &= &{(2\omega)^4\over \bar {\cal C}}
    {}_{2}C_\Lambda^{\rm trans} , \qquad
  {}_{-2}B_\Lambda^{\rm ref} 
  ={(2\omega)^4 \over \bar {\cal C}}  {}_{2}B_\Lambda^{\rm ref}.
 \label{relcoeff}
\end{eqnarray}
Using the above relations, the coefficients 
${\cal N}_s$ are rewritten as 
\begin{equation}
 {\cal N}^{\rm out}_s={1\over 2i\omega^3}|N_s^{\rm out}|^2 ,
\qquad
  {\cal N}^{\rm down}_s={1\over 2i\omega^2 k}|N_s^{\rm down}|^2 ,
\end{equation}
with
\begin{eqnarray}
|N_s^{{\rm out}}|^2 
&\equiv&
{2^{3s-2}\omega^{2s+2} \over |{\cal C}|^{s/2-1}}
{1\over |{}_sB^{\rm inc}_\Lambda|^{2}}, 
\label{eq:Namp-out} \\
|N_s^{{\rm down}}|^2 &\equiv&
{2^{-3s-2} k^{-2s+2} 
|{\cal C}|^{s/2+1}\over |\kappa_2|^{s/2-1}|\kappa_1|^{s}
(2Mr_+)^{2s-1}}
  {|{}_sB^{\rm trans}_\Lambda|^{2}\over 
   |{}_sB^{\rm inc}_\Lambda|^{2}\, 
   |{}_sC^{\rm trans}_\Lambda|^{2}}.
\label{eq:Namp-down}
\end{eqnarray}
Hence, we finally obtain  
\begin{eqnarray}
h_{\mu\nu}^{\rm rad} &= & \mu\int d\omega 
  \sum_{\ell m}{1\over 2i\omega^3}\Bigl(
   N^{\rm out}_s 
    {}_s\Pi^{\rm out}_{\Lambda,\mu\nu}(x) 
    \int {d\tau\over \Sigma}\, \bar \phi_{\Lambda}^{out}(\tau)
\cr &&\qquad\qquad
   + {\omega\over k}N^{\rm down}_s 
     {}_s\Pi^{\rm down}_{\Lambda,\mu\nu}(x) 
    \int {d\tau\over\Sigma}
      \, \bar \phi_{\Lambda}^{\rm down}(\tau)
    \Bigr)+({\rm c.c.}),
\label{hrad}
\end{eqnarray}
where 
\begin{equation}
\phi_\Lambda^{{\rm (out/down)}}(\tau) :=
N_s^{{\rm (out/down)}} \Sigma(z(\tau))
{}_s\Pi_{\Lambda,\gamma\delta}^{{\rm (out/down)}}(z(\tau))
u^\gamma(\tau) u^\delta(\tau), 
\end{equation}
whose extension to a field is 
$ \phi^{\rm (out/down)}_{\Lambda}(x)$ 
defined in Eq.~(\ref{eq:def-phi-out}).

\section{Mano-Suzuki-Takasugi method} \label{sec:MST}
Mano, Suzuki and Takasugi formulated a method of
constructing a homogeneous solution for the radial Teukolsky
equation in two kinds of series by using
the Coulomb wave function and the hypergeometric functions
\cite{Mano:1996vt,Mano:1996gn,Sasaki:2003xr}.
By applying this method under slow motion approximation,
we can express homogeneous solutions in an analytic form.
Furthermore, this method determines the asymptotic amplitudes 
of homogeneous solutions without numerical integration.
This allows us to compute the gravitational wave flux
at infinity and on the horizon with a high accuracy
\cite{Fujita:2004rb}.
We summarize this method in this appendix. 

\subsection{Outer solution of radial Teukolsky equation}
According to \citen{Mano:1996gn,Mano:1996vt,Sasaki:2003xr},
we can expand ${}_sR_C^{\nu}$, a homogeneous solution
of the radial Teukolsky equation (\ref{eq:radial-Teuk}),
in terms of the Coulomb wave functions as 
\begin{eqnarray}
{}_sR_C^{\nu} 
&=&
\frac{\Gamma(\nu+1-s+i\epsilon)}{\Gamma(2\nu+2)}
\hat{z}^{-s}(2\hat{z})^{\nu} e^{-i\hat{z}}
\left(1-\frac{\epsilon\kappa}{\hat{z}}\right)^{-s-i\epsilon_+}
\cr && \times
\sum_{n=-\infty}^{\infty} \!\! (-2i\hat{z})^n
\frac{(\nu+1+s-i\epsilon)_n}{(2\nu+2)_{2n}} a_n^{\nu,s}
\cr && \hspace*{1.5cm} \times
{}_1F_1(n+\nu+1-s+i\epsilon,2n+2\nu+2 ; 2i\hat{z}),
\label{eq:Coulomb-series}
\end{eqnarray}
where $\epsilon=2M\omega$, $\epsilon_+=\epsilon+\tau$,
$\tau=\kappa^{-1}(\epsilon-ma/M)$,
$\kappa=\sqrt{1-(a/M)^2}$, $(x)_n:=\Gamma(x+n)/\Gamma(x)$,
$\hat{z}:=\omega(r-r_-)$, and $r_-=M-\sqrt{M^2-a^2}$.
The coefficients $a_n^{\nu,s}$ satisfies the following 
three term recurrence relation, 
\begin{equation}
\alpha_n^{\nu} a_{n+1}^{\nu,s}+\beta_n^{\nu} a_n^{\nu,s}
+\gamma_n^{\nu} a_{n-1}^{\nu,s}=0,
\label{eq:recurrence}
\end{equation}
where
\begin{eqnarray}
\alpha_n^{\nu} &=& 
\frac{i\epsilon \kappa
(n+\nu+1+s+i\epsilon)(n+\nu+1+s-i\epsilon)(n+\nu+1+i\tau)}
{(n+\nu+1)(2n+2\nu+3)}, \nonumber \\
\beta_n^{\nu} &=& 
-\lambda-s(s+1)+(n+\nu)(n+\nu+1)
+\epsilon^2+\epsilon(\epsilon-mq) 
 +\frac{\epsilon (\epsilon-mq)(s^2+\epsilon^2)}{(n+\nu)(n+\nu+1)},
\nonumber \\
\gamma_n^{\nu} &=&
-\frac{i\epsilon \kappa (n+\nu-s+i\epsilon)
(n+\nu-s-i\epsilon)(n+\nu-i\tau)}
{(n+\nu)(2n+2\nu-1)},
\label{eq:coeff-Cwave-fnc}
\end{eqnarray}
and $q=a/M$. The renormalized angular momentum $\nu$ is determined
by the conditions 
\begin{equation}
\lim_{n\rightarrow\infty}
n\frac{a_n^{\nu,s}}{a_{n-1}^{\nu,s}} = \frac{i\epsilon\kappa}{2}, 
\qquad
\lim_{n\rightarrow -\infty}
n\frac{a_n^{\nu,s}}{a_{n+1}^{\nu,s}} = -\frac{i\epsilon\kappa}{2}.
\end{equation}
Under this condition, the series of Coulomb wave functions
(\ref{eq:Coulomb-series}) converges for any $r>r_+$.

From the equations in (\ref{eq:coeff-Cwave-fnc}),
we can show that $\alpha_{-n}^{-\nu-1}=\gamma_n^\nu$ and
$\beta_{-n}^{-\nu-1}=\beta_n^\nu$.
By using these relations, we can find that
$a_n^{-\nu-1,s}=a_{-n}^{\nu,s}$ and
\begin{equation}
\lim_{n\rightarrow\infty}
n\frac{a_n^{-\nu-1,s}}{a_{n-1}^{-\nu-1,s}} = \frac{i\epsilon\kappa}{2}, 
\qquad
\lim_{n\rightarrow -\infty}
n\frac{a_n^{-\nu-1,s}}{a_{n+1}^{-\nu-1,s}} = -\frac{i\epsilon\kappa}{2}.
\end{equation}
This fact shows that ${}_sR_C^{-\nu-1}$ is also a solution of the
radial Teukolsky equation, which converges within the region $r>r_+$.

\subsection{In-going and up-going solutions}
The in-going solution of the radial Teukolsky equation is given
in terms of the Coulomb type solutions (\ref{eq:Coulomb-series}) 
as 
\begin{equation}
_sR_\Lambda^{{\rm in}} =
A_s e^{i\epsilon\kappa}(
K_{s,\nu}~_sR_{C}^{\nu} + K_{s,-\nu-1}~_sR_{C}^{-\nu-1}),
\label{eq:Rin}
\end{equation}
where
\begin{eqnarray}
A_2 &=& \bar{{\cal C}}\left(\frac{\omega}{\epsilon\kappa}\right)^4
\frac{\Gamma(3-2i\epsilon_+)}{\Gamma(-1-2i\epsilon_+)}
\left| \frac{\Gamma(\nu-1+i\epsilon)}
{\Gamma(\nu+3+i\epsilon)}\right|^2, \quad A_{-2} = 1, \\
K_{s,\nu} &=&
\frac{(2\epsilon\kappa)^{s-\nu-r}2^{-s} i^r}
{(\nu+1+i\tau)_r(\nu+1+s+i\epsilon)_r}
\cr && \times
\frac{\Gamma(1-s-2i\epsilon_+)\Gamma(n+2\nu+2)\Gamma(n+2\nu+1)}
{\Gamma(r+\nu+1-s+i\epsilon)\Gamma(\nu+1-s-i\epsilon)
\Gamma(\nu+1-i\tau)} \cr
&& \times \left[\sum_{n=r}^{\infty} (-1)^n
\frac{(r+2\nu+1)_n(\nu+1+s+i\epsilon)_n(\nu+1+i\tau)_n}
{(n-r)!~(\nu+1-s-i\epsilon)_n(\nu+1-i\tau)_n}a_n^{\nu,s}\right] \cr
&& \times \left[\sum_{n=-\infty}^{r}
\frac{(-1)^n}{(r-n)!~(r+2\nu+2)_n}
\frac{(\nu+1+s-i\epsilon)_n}{(\nu+1-s+i\epsilon)_n}
a_n^{\nu,s}\right]^{-1}.
\end{eqnarray}
Here $r$ is an arbitrary integer and $K_{s,\nu}$ is independent
of the choice of $r$.

Next, we consider the up-going solution.
$_sR_C^{\nu}$ can be divided into two parts as
\begin{equation}
{}_sR_C^{\nu} = {}_sR_+^{\nu} + {}_sR_-^{\nu},
\end{equation}
where
\begin{eqnarray}
{}_sR_+^{\nu} &=&
e^{-\pi\epsilon}e^{i\pi(\nu+1-s)}e^{-i\hat{z}}
(2\hat{z})^{\nu}\hat{z}^{-s}
\left(1-\frac{\epsilon\kappa}{\hat{z}}\right)^{-s-i\epsilon_+}
\frac{\Gamma(\nu+1-s+i\epsilon)}{\Gamma(\nu+1+s-i\epsilon)} \cr
&& \times
\sum_{n=-\infty}^{\infty}(2i\hat{z})^n a_n^{\nu,s}
\Psi(n+\nu+1-s+i\epsilon,2n+2\nu+2; 2i\hat{z}), \\
{}_sR_-^{\nu} &=&
e^{-\pi\epsilon}e^{-i\pi(\nu+1+s)}e^{i\hat{z}}
(2\hat{z})^{\nu}\hat{z}^{-s}
\left(1-\frac{\epsilon\kappa}{\hat{z}}\right)^{-s-i\epsilon_+}
\cr && \times
\sum_{n=-\infty}^{\infty} \!\! (2i\hat{z})^n
\frac{(\nu+1+s-i\epsilon)_n}{(\nu+1-s+i\epsilon)_n}a_n^{\nu,s}
\cr && \hspace*{1.5cm} \times
\Psi(n+\nu+1+s-i\epsilon,2n+2\nu+2; -2i\hat{z}), 
\end{eqnarray}
and $\Psi(a,c;x)$ 
is the irregular confluent hypergeometric function. 
From the asymptotic form of $\Psi(a,c;x)$,  
\begin{equation}
\Psi(a,c; x) \rightarrow x^{-a}, \qquad
(|x|\rightarrow\infty),
\end{equation}
the asymptotic forms of $_sR_+^{\nu}$ and $_sR_-^{\nu}$ 
become 
\begin{equation}
_sR_+^{\nu} = ~_sA_+^{\nu}z^{-1}e^{-i(z+\epsilon\ln z)},
\qquad
_sR_-^{\nu} = ~_sA_-^{\nu}z^{-1-2s}e^{i(z+\epsilon\ln z)},
\end{equation}
where
\begin{eqnarray}
{}_sA_+^{\nu} &=& 
e^{-\pi\epsilon/2}e^{i\pi(\nu+1-s)/2}2^{s-1-i\epsilon}
\frac{\Gamma(\nu+1-s+i\epsilon)}{\Gamma(\nu+1+s-i\epsilon)}
\sum_{n=-\infty}^{\infty} \!\! a_n^{\nu,s}, \\
{}_sA_-^{\nu} &=&
e^{-\pi\epsilon/2}e^{-i\pi(\nu+1+s)/2}2^{-s-1+i\epsilon}
\sum_{n=-\infty}^{\infty} \!\! (-1)^n
\frac{(\nu+1+s-i\epsilon)_n}{(\nu+1-s+i\epsilon)_n}
a_n^{\nu,s}.
\end{eqnarray}
This shows that $_sR_-^{\nu}$ ($_sR_+^{\nu}$) satisfies the
up-going (down-coming) boundary condition at infinity.
So we can take the up-going solution as
\begin{equation}
{}_sR_\Lambda^{{\rm up}} = B_s {}_sR_-^{\nu},
\label{eq:Rup}
\end{equation}
where $B_2=\bar{C}\omega^{2s}$ and $B_{-2}=1$.
Taking the limit $r^*\to\pm\infty$ 
in Eqs.~(\ref{eq:Rin}) and (\ref{eq:Rup}) 
by means of the asymptotic form of $r^*$, 
\begin{eqnarray}
\omega r^* &\rightarrow&
\hat{z}+\epsilon\ln\hat{z}-\epsilon\ln\epsilon \quad
(r\rightarrow\infty), \\
kr^* &\rightarrow&
\epsilon_+\ln(-x) + \kappa\epsilon_+
+\frac{2\kappa\epsilon_+}{1+\kappa}\ln\kappa \quad
(r\rightarrow r_+),
\end{eqnarray}
we find that the coefficients which appear in the asymptotic
forms of Eqs.~(\ref{eq:asymptotic-in}) and (\ref{eq:asymptotic-up})
are given by
\begin{eqnarray}
{}_sB_\Lambda^{{\rm inc}} &=&
\frac{A_s e^{i\epsilon\kappa}}{\omega}
\left[ K_{s,\nu} - ie^{-i\pi\nu}
\frac{\sin\pi(\nu-s+i\epsilon)}{\sin\pi(\nu+s-i\epsilon)}
K_{s,-\nu-1} \right] {}_sA_+^{\nu}, \\
{}_sB_\Lambda^{{\rm trans}} &=&
A_s\left(\frac{\epsilon\kappa}{\omega}\right)^{2s}
\sum_{n=-\infty}^{\infty} a_n^{\nu,s}, \\
{}_sC_\Lambda^{{\rm trans}} &=&
\omega^{-1-2s}e^{i\epsilon\ln\epsilon} {}_sA_-^{\nu}. 
\end{eqnarray}

\section{Spheroidal harmonics} \label{sec:spheroidal}
Here, we review the formalism to represent the spin-weighted spheroidal 
harmonics in a series of Jacobi polynomials based on
Ref.~\citen{Fackerell}, which was slightly improved in 
Ref.~\citen{Fujita:2004rb}.

We first transform the angular part of the Teukolsky equation 
(\ref{eq:spheroid-eq}) as 
\begin{eqnarray}
\bigg[(1-x^{2})\frac{d^2}{dx^{2}}-2x\frac{d}{dx}+{\varpitmp}^{2} x^{2}
\hspace*{4cm} && \cr
-\frac{m^{2}+s^{2}+2msx}{1-x^{2}}-2s\varpitmp x+{}_sE_{\ell m}(\varpitmp)
\bigg]
{}_sS_{\ell m}^{\varpitmp}(x) &=& 0 \,,
\label{eq:Sphe diff}
\end{eqnarray}
where $\varpitmp = a\,\omega, x = \cos\theta$
and ${}_sE_{\ell m}(\varpitmp)=\lambda+s(s+1)-\varpitmp^{2}+2m\varpitmp$.
The angular function $_{s}S_{\ell m}^{\varpitmp}(x)$ is called the
spin-weighted spheroidal harmonics. Equation (\ref{eq:Sphe diff}) 
is a Sturm-Liouville type eigenvalue equation with
regular boundary conditions at $x=\pm 1$. 
Since there are a countable number of eigenvalues 
for fixed parameters $s$, $m$ and $\varpitmp$, 
we introduced an index $\ell$ starting with max($|m|,|s|$) 
as such a label 
that sorts the eigenvalues ${}_sE_{\ell m}(\varpitmp)$ in an ascending 
order. 
When $\varpitmp=0$, $_{s}S_{\ell m}^{\varpitmp}(x)$ 
is reduced to the spin-weighted spherical 
harmonics, and the eigenvalue ${}_sE_{\ell m}(\varpitmp)$ becomes $\ell(\ell+1)$.
We normalize the amplitude of $_{s}S_{\ell m}^{\varpitmp}(x)$ as 
\begin{eqnarray}
\label{eq:normalSp}
\int _{0}^{\pi}\left |{}_{s}S_{\ell m}^{\varpitmp}\right |^2\sin \theta d\theta=1 \,.
\end{eqnarray}

The differential equation (\ref{eq:Sphe diff}) has singularities at
$x=\pm 1$ and at $x=\infty$. We transform the angular function as 
\begin{eqnarray}
_{s}S_{\ell m}^{\varpitmp}(x) \equiv
e^{\varpitmp x}\left(\frac{1-x}{2}\right)^{\frac{\alpha}{2}}
\left(\frac{1+x}{2}\right)^{\frac{\beta}{2}}\, _{s}U_{\ell m}(x) \,,
\label{eq:SpheU}
\end{eqnarray}
and 
\begin{eqnarray}
\label{eq:SpheV}
_{s}S_{\ell m}^{\varpitmp}(x) \equiv
e^{-\varpitmp x}\left(\frac{1-x}{2}\right)^{\frac{\alpha}{2}}
\left(\frac{1+x}{2}\right)^{\frac{\beta}{2}}\, _{s}V_{\ell m}(x) \,,
\label{Vseries}
\end{eqnarray}
where $\alpha = |m+s|$ and $\beta = |m-s|$. Then, 
Eq.~(\ref{eq:Sphe diff}) becomes
\begin{eqnarray}
\label{eq:proto-Jacobi}
&&(1-x^{2})\,_{s}U_{\ell m}''(x)+\left[\beta-\alpha-(2+\alpha+\beta)x\right]\,_{s}U_{\ell m}'(x)
\nonumber\\
&&\quad
+\left[\,_{s}E_{\ell m}(\varpitmp)-\frac{\alpha+\beta}{2}\left(\frac{\alpha+\beta}{2}+
1\right)\right]\,_{s}U_{\ell m}(x)\nonumber \\
&&\quad
=\varpitmp\left[-2(1-x^{2})\,_{s}U_{\ell m}'(x)+(\alpha+\beta+2s+2)x\,_{s}U_{\ell m}(x)
\right.
\nonumber\\
&&\quad\quad\left. 
-(\varpitmp+\beta-\alpha)\,_{s}U_{\ell m}(x)\right] \,,
\end{eqnarray}
and 
\begin{eqnarray}
\label{eq:proto-JacobiV}
&&(1-x^{2})\,_{s}V_{\ell m}''(x)+\left[\beta-\alpha-(2+\alpha+\beta)x\right]\,
_{s}V_{\ell m}'(x)
\nonumber\\
&&\quad
+\left[\,_{s}E_{\ell m}(\varpitmp)-\frac{\alpha+\beta}{2}\left(\frac{\alpha+\beta}{2}+
1\right)\right]\,_{s}V_{\ell m}(x)\nonumber \\
&=&\varpitmp\left[2(1-x^{2})\,_{s}V_{\ell m}'(x)-(\alpha+\beta-2s+2)x\,_{s}V_{\ell m}(x)
\right.
\nonumber\\
&&\quad
\left.
-(\varpitmp-\beta+\alpha)\,_{s}V_{\ell m}(x)\right] \,.
\end{eqnarray}
From Eqs.~(\ref{eq:SpheU}) and (\ref{eq:SpheV}), we find
\begin{eqnarray}
\label{eq:sphUtoV}
\,_{s}V_{\ell m}(x)={\rm exp}(2\varpitmp x)\,_{s}U_{\ell m}(x) \,.
\end{eqnarray}

When $\varpitmp=0$, the right-hand sides of Eqs.~(\ref{eq:proto-Jacobi})
and (\ref{eq:proto-JacobiV}) are zero, and they reduce to
the differential equation satisfied by the Jacobi polynomials, 
\begin{eqnarray}
&&(1-x^{2})\,P_{n}^{(\alpha,\beta)}{}^{''}(x)
+\left[\beta-\alpha-(2+\alpha+\beta)x\right]\,P_{n}^{(\alpha,\beta)}{}^{'}(x)
\nonumber\\
&&
\quad +n(n+\alpha+\beta+1)\,P_{n}^{(\alpha,\beta)}(x)=0.
\label{eq:Jacobi}
\end{eqnarray}
In this limit, the eigenvalue $_{s}E_{\ell m}(\varpitmp)$ in the equation 
(\ref{eq:proto-Jacobi}) becomes $\ell(\ell+1)$, 
where $n=\ell-(\alpha+\beta)/2=\ell-{\rm max}(\mid m\mid ,\mid s\mid )$.
Here, the Jacobi polynomials are defined by the Rodrigue's formula by
\begin{eqnarray}
P_{n}^{(\alpha,\beta)}(x) :=
\frac{(-1)^{n}}{2^{n}\,n!}(1-x)^{-\alpha}(1+x)^
{-\beta}\left(\frac{d}{dx}\right)^{n}\left[(1-x)^{\alpha+n}(1+x)^{\beta+n}
\right].
\end{eqnarray}

Now, we expand $_{s}U_{\ell m}(x)$ and $_{s}V_{\ell m}(x)$ in a series of
Jacobi polynomials: 
\begin{eqnarray}
\label{eq:Jacobi-series}
_{s}U_{\ell m}(x)&=&\sum_{n=0}^{\infty}\,_{s}A_{\ell m}^{(n)}(\varpitmp)\,P_{n}^{(\alpha,\beta)}(x) \,,
\\ 
\label{eq:Jacobi-series2}
_{s}V_{\ell m}(x)&=&\sum_{n=0}^{\infty}\,_{s}B_{\ell m}^{(n)} \,P_{n}^{(\alpha,\beta)}(x) \,.
\end{eqnarray}
The expansion coefficients $_{s}A_{\ell m}^{(n)}(\varpitmp)$ 
and $_{s}B_{\ell m}^{(n)}(\varpitmp)$  satisfy the recurrence 
relations
\begin{eqnarray}
\alpha^{(0)}\,_{s}A_{\ell m}^{(1)}(\varpitmp)
+\beta^{(0)}\,_{s}A_{\ell m}^{(0)}(\varpitmp)&=&0, 
\label{eq:3termElm0}\\
\alpha^{(n)}\,_{s}A_{\ell m}^{(n+1)}(\varpitmp)
+\beta^{(n)}\,_{s}A_{\ell m}^{(n)}(\varpitmp)
+\gamma^{(n)}\,_{s}A_{\ell m}^{(n-1)}(\varpitmp)&=&0, \, (n\ge 1) \,,
\label{eq:3termElm}
\end{eqnarray}
with
\begin{eqnarray}
\alpha^{(n)}&:=&
\frac{4\varpitmp(n+\alpha+1)(n+\beta+1)(n+(\alpha+\beta)/2+1-s)}
{(2n+\alpha+\beta+2)(2n+\alpha+\beta+3)},\nonumber \\
\beta^{(n)}&:=&
\,_{s}E_{\ell m}(\varpitmp)+\varpitmp ^2-\left(n+\frac{\alpha+\beta}{2}\right)
\left(n+\frac{\alpha+\beta}{2}+1\right)\nonumber \\
&&+\frac{2\varpitmp s(\alpha-\beta)(\alpha+\beta)}
{(2n+\alpha+\beta)(2n+\alpha+\beta+2)},\nonumber \\
\gamma^{(n)}&:=&
-\frac{4\varpitmp n(n+\alpha+\beta)(n+(\alpha+\beta)/2+s)}
{(2n+\alpha+\beta-1)(2n+\alpha+\beta)} \,,
\end{eqnarray}
and 
\begin{eqnarray}
\tilde{\alpha}^{(0)}\,_{s}B_{\ell m}^{(1)}(\varpitmp)
+\tilde{\beta}^{(0)}\,_{s}B_{\ell m}^{(0)}(\varpitmp)
&=&0, \nonumber \\
\tilde{\alpha}^{(n)}\,_{s}B_{\ell m}^{(n+1)}(\varpitmp)
+\tilde{\beta}^{(n)}\,_{s}B_{\ell m}^{(n)}(\varpitmp)
+\tilde{\gamma}^{(n)}\,_{s}B_{\ell m}^{(n-1)}(\varpitmp)&=&0, \quad (n\ge 1) \,,
\label{eq:3termElm2}
\end{eqnarray}
with
\begin{eqnarray}
\tilde{\alpha}^{(n)}&:=&
-\frac{4\varpitmp(n+\alpha+1)(n+\beta+1)(n+(\alpha+\beta)/2+1+s)}
{(2n+\alpha+\beta+2)(2n+\alpha+\beta+3)},\nonumber \\
\tilde{\beta}^{(n)}&:=&
\,_{s}E_{\ell m}(\varpitmp)+\varpitmp ^2-\left(n+\frac{\alpha+\beta}{2}\right)
\left(n+\frac{\alpha+\beta}{2}+1\right)\nonumber \\
&&+\frac{2\varpitmp s(\alpha-\beta)(\alpha+\beta)}{(2n+\alpha+\beta)(2n+\alpha+\beta+2)},\nonumber \\
\tilde{\gamma}^{(n)}&:=&
\frac{4\varpitmp n(n+\alpha+\beta)(n+(\alpha+\beta)/2-s)}
{(2n+\alpha+\beta-1)(2n+\alpha+\beta)} \,.
\end{eqnarray}

The eigenvalues $\,{_s}E_{\ell m}(\varpitmp)$ are determined 
in a way similar to 
the renormalized angular momentum $\nu$. 
The three-term recurrence relation Eq.~(\ref{eq:3termElm}) 
has two independent solutions, which respectively behave for large $n$ as
\begin{eqnarray}
&&
A_{(1)}^{(n)}\sim \frac{({\rm const.})\,(-\varpitmp)^n}
{\Gamma(n+(\alpha+\beta+3)/2-s)} \,, \label{eq:AlmMin}
\\ &&
A_{(2)}^{(n)}\sim ({\rm const.})\,\varpitmp^n \Gamma(n+(\alpha+\beta+1)/2+s)\,.
\label{eq:AlmDom}
\end{eqnarray}
The first one, $A_{(1)}^{(n)}$, is the minimal solution, and the 
second one, $A_{(2)}^{(n)}$, is a dominant solution, since
$\displaystyle\lim_{n\rightarrow \infty}A_{(1)}^{(n)}/A_{(2)}^{(n)}=0$. 
In the case of the dominant solution
these coefficients $A_{(2)}^{(n)}$ increase with $n$, 
and the series~(\ref{eq:Jacobi-series}) diverges for 
all values of $x$. 
In the case of the minimal solution 
this series converges. 
Thus, we have to choose $A_{(1)}^{(n)}$ in the 
series expansion~(\ref{eq:Jacobi-series}). 
For a general ${_s}E_{\ell m}(\varpitmp)$, 
$A_{(1)}^{n}$ 
does not satsify the relation (\ref{eq:3termElm0}). Hence, 
the requirement to satisfy this condition determines 
the descrete eigen values ${_s}E_{\ell m}(\varpitmp)$.

{}As a practical way to obtain $A_{(1)}^{(n)}$ 
as well as ${_s}E_{\ell m}(\varpitmp)$, 
we introduce 
\begin{equation}
R_n\equiv {A_{(1)}^{n}\over A_{(1)}^{n-1}},\qquad
L_n\equiv {A_{(1)}^{n}\over A_{(1)}^{n+1}}.
\end{equation}
The ratio $R_n$ can be expressed as a continued fraction,
\begin{equation}
R_n =-{\gamma^{(n)}\over {\beta^{(n)}+\alpha^{(n)} R_{n+1}}}
=-{\gamma^{(n)}\over \beta^{(n)}-}
{\alpha^{(n)}\gamma^{(n+1)}\over \beta^{(n+1)}-}
{\alpha^{(n+1)}\gamma^{(n+2)}\over \beta^{(n+2)}-}\cdots . 
\label{eq:RncontElm}
\end{equation}
We can also express $L_n$ in a similar way as
\begin{eqnarray}
L_n&=&-{\alpha^{(n)}\over {\beta^{(n)}+\gamma^{(n)} L_{n-1}}}
\cr
&=&-{\alpha^{(n)}\over \beta^{(n)}-}\,
{\alpha^{(n-1)}\gamma^{(n)}\over \beta^{(n-1)}-}\,
{\alpha^{(n-2)}\gamma^{(n-1)}\over \beta^{(n-2)}-}\cdots
{\alpha^{(1)}\gamma^{(2)}\over \beta^{(1)}-}\,
{\alpha^{(0)}\gamma^{(1)}\over \beta^{(0)}}.
\label{eq:LncontElm}
\end{eqnarray}
This expressions for $R_n$ and $L_n$ are valid 
if the continued fraction~(\ref{eq:RncontElm}) converge. 
(Notice that the last step of Eq.~(\ref{eq:LncontElm})
is not a continued fraction, but just a rational function.)
By using the properties of the three-term recurrence relations, 
it is proved that the continued fraction~(\ref{eq:RncontElm})
converges as long as the eigenvalue $_{s}E_{\ell m}(\varpitmp)$
is finite. 

Dividing Eq.~(\ref{eq:3termElm}) by the expansion coefficients 
$_{s}A_{\ell m}^{(n)}$, we obtain 
\begin{eqnarray}
\beta^{(n)}+\alpha^{(n)}R_{n+1}+\gamma^{(n)}L_{n-1}=0 \,.
\label{eq:determine_elm}
\end{eqnarray}
We replace $R_{n+1}$ and $L_{n-1}$ by Eqs.~(\ref{eq:RncontElm})
and (\ref{eq:LncontElm}). 
Then we can determine the eigenvalue $\,_{s}E_{\ell m}$ as a root of 
Eq.~(\ref{eq:determine_elm}). There are many roots, and the above 
equations for all value of $n$ are equivalent. 
In practice, however, 
we truncate the continued fractions at finite lengths. 
In this case the most efficient way is to choose the equation with 
$n=n_\ell:=\ell-(\alpha+\beta)/2$. With this choice all terms in  
Eq.~(\ref{eq:determine_elm}) become $O(\varpitmp^2)$, and 
the length of the continued fractions that we must keep  
to achieve a given accuracy goal is the shortest.

As was done in Fujita and Tagoshi's paper, in general, we can 
adopt {\rm Brent's algorithm}\cite{Recipes} 
in order to determine $_{s}E_{\ell m}(\varpitmp)$. 
However, when $|\varpitmp|$ is not large, 
we can derive an analytic expression
for $_{s}E_{\ell m}(\varpitmp)$. The result is 
\begin{eqnarray}
_{s}E_{\ell m}(\varpitmp) 
    = \ell (\ell +1) -\frac{2 s^2 m}{\ell (\ell +1)} \varpitmp 
	+ \left[H(\ell+1)-H(\ell)-1\right]\varpitmp^2 +O(\varpitmp^3),
\end{eqnarray}
with
\begin{eqnarray}
H(\ell)=\frac{2(\ell^2-m^2)(\ell^2-s^2)^2}{(2\ell-1)\ell^3(2\ell+1)}.
\end{eqnarray}

After we obtain the eigenvalues $_{s}E_{\ell m}(\varpitmp)$, 
we can easily determine all the coefficients. 
The coefficient with $n=n_\ell$
is usually the largest term. The ratio of the other terms 
to the dominant term, i.e.
$A_{(1)}^{(n)}/A_{(1)}^{(n_{\ell})}$, 
can be determined 
in the most efficient way with a minimal error due to truncation 
using Eqs.~(\ref{eq:RncontElm}) and (\ref{eq:LncontElm})
for $0<n<n_\ell$ and $n>n_\ell$, respectively.  

The coefficient of the leading term 
$A_{(1)}^{(n_{\ell})}$ $\big({}_{s}A_{\ell m}^{(n_{\ell})}\big)$ 
is determined by the normalization condition. 
Since (\ref{Vseries}) represents the same eigen function, 
the series~(\ref{eq:Jacobi-series2}) should converge 
for the same eigen values $_{s}E_{\ell m}(\varpitmp)$,
constituting the minimal solution of the recurrence relation
Eq.~(\ref{eq:3termElm2}).
As in the case of $\{A_{(1)}^{(n)}\}$,  
we have 
\begin{eqnarray}
\frac{B_{(1)}^{(n)}}{B_{(1)}^{(n-1)}}&=&
-{\tilde{\gamma}^{(n)}\over \tilde{\beta}^{(n)}-}\,
{\tilde{\alpha}^{(n)}\tilde{\gamma}^{(n+1)}\over \tilde{\beta}^{(n+1)}-}\,
{\tilde{\alpha}^{(n+1)}\tilde{\gamma}^{(n+2)}\over \tilde{\beta}^{(n+2)}-}\cdots , 
\label{eq:RncontElm2}
\\
\frac{B_{(1)}^{(n)}}{B_{(1)}^{(n+1)}}&=&
-{\tilde{\alpha}^{(n)}\over \tilde{\beta}^{(n)}-}\,
{\tilde{\alpha}^{(n-1)}\tilde{\gamma}^{(n)}\over \tilde{\beta}^{(n-1)}-}\,
{\tilde{\alpha}^{(n-2)}\tilde{\gamma}^{(n-1)}\over \tilde{\beta}^{(n-2)}-}\cdots
{\tilde{\alpha}^{(1)}\tilde{\gamma}^{(2)}\over \tilde{\beta}^{(1)}-}\,
{\tilde{\alpha}^{(0)}\tilde{\gamma}^{(1)}\over \tilde{\beta}^{(0)}}.
\label{eq:LncontElm2}
\end{eqnarray}
From these equations, we can determine the ratios of 
all coefficients, $B_{(1)}^{(n)}/B_{(1)}^{(n_{\ell})}$. 

Now, we determine the values of the two coefficients
$A_{(1)}^{(n_{\ell})}$ and $B_{(1)}^{(n_{\ell})}$ that 
determines the overall normalization.
Since Eq.~(\ref{eq:sphUtoV}) must hold for any value of $x$, 
we can set $x=1$ in it to obtain
\begin{eqnarray}
&& \hspace*{-1cm}
{}_{s}B_{\ell m}^{(n_{\ell})}(\varpitmp)\sum_{n=0}^{\infty}
\frac{\,_{s}B_{\ell m}^{(n)}(\varpitmp)}
{\,_{s}B_{\ell m}^{(n_{\ell})}(\varpitmp)}
\frac{\Gamma(n+\alpha+1)}{\Gamma(n+1)\,\Gamma(\alpha+1)}
\nonumber \\
&& \hspace*{5mm} =
{\rm exp}(2\varpitmp)\,_{s}A_{\ell m}^{(n_{\ell})}(\varpitmp)\sum_{n=0}^{\infty}
\frac{\,_{s}A_{\ell m}^{(n)}(\varpitmp)}{\,_{s}A_{\ell m}^{(n_{\ell})}(\varpitmp)}
\frac{\Gamma(n+\alpha+1)}{\Gamma(n+1)\,\Gamma(\alpha+1)}.
\label{eq:normalization1}
\end{eqnarray}
On the other hand, 
from the normalization condition~(\ref{eq:normalSp}), we find
\begin{eqnarray}
\int_{-1}^{1} \!\! dx
\bigg(\frac{1-x}{2}\bigg)^{\alpha}
\bigg(\frac{1+x}{2}\bigg)^{\beta} \!
\sum_{n_{1}=0}^{\infty} \! {}_{s}A_{\ell m}^{(n_{1})}
P_{n_{1}}^{(\alpha,\beta)}(x) \!
\sum_{n_{2}=0}^{\infty} \! {}_{s}B_{\ell m}^{(n_{2})}
P_{n_{2}}^{(\alpha,\beta)}(x)=1.
\label{eq:AlmBlm}
\end{eqnarray}
Because the Jacobi polynomials are orthogonal, we have 
\begin{eqnarray}
&& \hspace*{-1cm}
\int_{-1}^{1}{\rm d}x\left(
\frac{1-x}{2}\right)^{\alpha}\left(\frac{1+x}{2}\right)^{\beta}
P_{n_{1}}^{(\alpha,\beta)}(x)P_{n_{2}}^{(\alpha,\beta)}(x)
\nonumber\\
&&\quad
=\frac{2\, \Gamma(n+\alpha+1)\Gamma(n+\beta+1)
\delta_{n_{1},n_{2}}}{(2n+\alpha+\beta+1)
\Gamma(n+1)\Gamma(n+\alpha+\beta+1)}.
\end{eqnarray}
Then, Eq.~(\ref{eq:AlmBlm}) reduces to 
\begin{eqnarray}
&&\sum_{n=0}^{\infty}\left[
\frac{\,_{s}A_{\ell m}^{(n)}}{\,_{s}A_{\ell m}^{(n_{\ell})}}\right]
\left[\frac{\,_{s}B_{\ell m}^{(n)}}{\,_{s}B_{\ell m}^{(n_{\ell})}}\right]
\frac{2\, \Gamma(n+\alpha+1)\Gamma(n+\beta+1)}
{(2n+\alpha+\beta+1)\Gamma(n+1)
\Gamma(n+\alpha+\beta+1)}
\nonumber\\
&&\quad
=\frac{1}{\,_{s}A_{\ell m}^{(n_{\ell})}\,_{s}B_{\ell m}^{(n_{\ell})}}.
\label{eq:normalization2}
\end{eqnarray}
Combining Eqs.~(\ref{eq:normalization1}) and (\ref{eq:normalization2}), 
we can determine the squares of $\,_{s}A_{\ell m}^{(n_{\ell})}$
and $\,_{s}B_{\ell m}^{(n_{\ell})}$.
Finally, we fix the signatures of $\,_{s}A_{\ell m}^{(n_{\ell})}$ and
$\,_{s}B_{\ell m}^{(n_{\ell})}$ so that 
$_{s}S_{\ell m}^{\varpitmp}(x)$ reduces to 
the spin-weighted spherical harmonics 
in the limit $\varpitmp\rightarrow 0$.

\end{appendix}

\end{document}